\documentclass[aps,prd,preprint,groupedaddress,showpacs,showkeys,fleqn,11pt]{revtex4-1}
\usepackage{graphicx}
\usepackage{amsmath}
\usepackage{array}
\usepackage{longtable}
\usepackage{natbib}
\begin{document}

\title{Self-gravitating Newtonian models of fermions with anisotropy and cutoff energy in their distribution function.}
\author{Marco Merafina}
\email{marco.merafina@roma1.infn.it}
\affiliation{Department of Physics, University of Rome La Sapienza, Piazzale Aldo Moro 2, I-00185 Rome, Italy}
\author{Giuseppe Alberti}
\email{alberti.signorkimi@libero.it}
\affiliation{Department of Physics, University of Rome La Sapienza, Piazzale Aldo Moro 2, I-00185 Rome, Italy}

\begin{abstract}
Systems of self-gravitating fermions constitute a topic of great interest in astrophysics, due to the wide field of applications. In this paper, we consider the gravitational equilibrium of spherically symmetric Newtonian models of collisionless semidegenerate fermions. We construct numerical solutions by taking into account the effects of the anisotropy in the distribution function and considering the prevalence of tangential velocity. In this way, our models generalize the solutions obtained for isotropic Fermi-Dirac statistics. We also extend the analysis to equilibrium configurations in the classical regime and in the fully degenerate limit, recovering, for different levels of anisotropy, hollow equilibrium configurations obtained in the Maxwellian regime. Moreover, in the limit of full degeneracy, we find a direct expression relating the anisotropy with the mass of the particles composing the system.
\end{abstract}

\pacs{98.10.+z -- 98.62.Gq}
\keywords{galactic halos -- anisotropy -- dark matter}
\maketitle

\section{Introduction}

The first evidence of dark matter was found by Zwicky \cite{zwicky33}, who noted (in studying the Coma cluster of galaxies) that the dispersion in the radial velocity of the galaxies was very large (around 1000 km s$^{-1}$) and that the stars and gas visible within the galaxies did not provide enough gravitational attraction to keep bound the cluster. In order to maintain the galaxies in the Coma cluster Zwicky concluded that the cluster had to contain a large amount of ``dunkle materie," i.e., ``dark matter." A confirmation of Zwicky's suggestion was furnished about 40 years later from optical and 21-cm observations of rotation curves of spiral galaxies that did not show the Keplerian falloff, implying the presence of an additional mass component and suggesting that spiral galaxies have a massive dark halo extending to several times the radius of the luminous matter and containing most of the total of their mass \cite{1987bt_book}.

Then, from the rotation curve it is possible to parametrize the distribution of matter within the halo. As a first approximation, we may schematize the halo as a spherical one and neglect the gravitational contribute of the disk. If we furthermore assume a rotation curve with constant velocity $v_{c}$, the corresponding dark halo density profile corresponds to $\rho(r) \sim 1/r^{2}$. The \emph{N}-body simulations are a powerful tool to study the behavior of particles forming dark halos and give us a better parametrization of the halo density, that we may write as
\begin{equation} \label{n_body}
\rho(r) = \frac{\rho_{0}}{{(\frac{r}{r_{s}})}^\lambda {[1 + {(\frac{r}{r_{s}})}^\mu]}^\nu}\ ,
\end{equation}
where $\lambda$, $\mu$ and $\nu$ are parameters defining the profile, while $r_{s}$ is the radius scale (the scale in which the profile changes shape, connected with the core radius). For $\lambda$ = 1, $\mu$ = 1, and $\nu$ = 3, Eq.(\ref{n_body}) corresponds to the Hernquist model \cite{1990hernquist} while the model proposed by Navarro, Frenk and White \cite{nfw96,nfw97} has $\lambda$ = 1, $\mu$ = 1, and $\nu$ = 2 and Moore \emph{et al.} \cite{mgqsl98,mgqsl99} proposed the model with $\lambda$ = 1.5, $\mu$ = 1, and $\nu$ = 1. These proposals have an unpleasant feature, due to the presence of a cusp when the density is evaluated at the center: to solve this problem, Burkert \cite{burkert_profile} has proposed a different expression for the density profile of dark matter
\begin{equation} \label{burkert}
\rho(r) = \frac{\rho_{0} r_{0}^{3}}{(r + r_{0}) (r^{2} + r_{0}^{2})}\ ,
\end{equation}
where the parameters $\rho_{0}$ and $r_{0}$ are free and represent, respectively, the central density and the radius scale, in the same sense of Eq.(\ref{n_body}). Equation (\ref{burkert}) is similar to the isothermal profile in the limit $r \ll r_{0}$ (\cite{burkert_profile} and references therein) and predicts a finite central density.

In order to find candidates for particles forming halos, many hypotheses have been advanced. In the field of cosmology \cite{2009rich_book}, the theory of structure formation assumes that dark matter is composed by massive particles weakly interacting with matter, named WIMPs (predicted, moreover, by supersymmetric extensions of the standard model of particle physics). Anyway, there are other candidates to explain the formation of dark matter, like axions \cite{1988cheng} - invented to prevent \emph{CP} violation in the strong interactions, and massive compact Halo objects (MACHOs), that represent the simplest way to hide baryons and detectable from their gravitational lensing \cite{1986pacz} and cold gas \cite{1997cpf}, an alternative to MACHOs (in this case, baryons are hidden in small clouds comprised of primordial helium and molecular hydrogen).

There is indeed another way to study the properties of dark halos, by considering a different approach which takes into account statistical mechanics. A particular interest is due to systems made by massive neutrinos, especially in connection with the problem of the formation and stability of large scale structures like galaxies and clusters of galaxies \cite{cmcc72,cmcc73}. This case represents the starting point of the discussion of dark matter in terms of fermions. If we generally think about a system of fermions (assumed to be collisionless), it is not possible to use Fermi-Dirac distribution without considering a cutoff in kinetic energy, because masses and radii of definite configurations are not finite. For this reason arises the necessity to introduce a cutoff energy in Fermi-Dirac statistics \cite{1983rufs}:
\begin{equation} \label{fd_cutoff}
\begin{cases}
f=\frac{g}{h^{3}} \frac{1 - e^{(\epsilon - \epsilon_{c})/kT}}{e^{(\epsilon - \mu)/kT} + 1} & \text{for $\epsilon\leq\epsilon_{c}$} \\
f=0 & \text{for $\epsilon > \epsilon_{c}$}\ ,
\end{cases}
\end{equation}
where $\epsilon$ is the kinetic energy of a single particle and $\epsilon_{c}$ the cutoff energy, $g = 2s + 1$ is the multiplicity of quantum states, $\mu$ is the chemical potential and $T$ is the temperature; in the limit $\epsilon_c\rightarrow\infty$ we recover the semidegenerate Fermi-Dirac distribution. In the fully degenerate case, we have a natural cutoff velocity associated to the Fermi energy. Then, the question of the relation between the Fermi energy $\epsilon_F$ and the cutoff energy $\epsilon_c$ is arising in order to better understand how the particles are distributed when $T\rightarrow 0$. We know that the energy variation due to a single particle (at constant entropy $S$ and volume $V$) cannot be larger than the maximum energy that a particle can have. This condition can be expressed by the inequality $(\partial U/\partial N)_{S,V}\leq\epsilon_c$, with $\epsilon_c$ being the maximum energy for a single particle. By the definition of the chemical potential $\mu = (\partial U/\partial N)_{S,V}$ we get the important relation $\mu\leq\epsilon_{c}$, valid at every regime and trivially verified in the nonquantum limit where $\mu<0$. In the fully degenerate case, indeed, from the condition $\epsilon_F =\mu$ we obtain that $\epsilon_F\leq\epsilon_c$: namely the cutoff energy never affects the value of the Fermi energy. Moreover, in the interval between $\epsilon_F$ and $\epsilon_c$ the density in phase space is zero, leading to the conclusion, valid only in the fully degenerate case, that the cutoff energy must coincide with the Fermi one. Therefore, the relation must be $\epsilon_F =\epsilon_c$. 

A system described by Eq.(\ref{fd_cutoff}) is characterized by an isotropic distribution of velocities of particles. Nevertheless, the presence of anisotropies in the momentum distribution, depending on the angular momentum (of particles forming the halos) $L = mrv_{t}$ ($m$, $r$, and $v_{t}$ are,  respectively, mass, radius, and tangential velocity), was found by \emph{N}-body simulations \cite{nav_2010} and observational data \cite{hhpmekv}. However, anisotropy is not a novelty in astrophysics. Michie \cite{michie61,michie63}, for example, studied the consequences of anisotropy for what concerns stellar systems, and, successively, many authors advanced proposals in this sense by considering stellar systems described by an anisotropic Boltzmann-like distribution function (\cite{bkmv09} and references therein). Anisotropic models had been also advanced for stellar models, for neutron stars (e.g., \cite{hh75}) or by considering generalized polytropic models (e.g., \cite{herr_barreto13a}). More generally, Lingam and Nguyen \cite{linguyen13} found a method consisting of two different approaches for deriving anisotropic distribution functions and they applied it to Veltmann, Hernquist, and Plummer models. Furthermore, a discussion about the effects of the presence of a local anisotropy in compact objects is in Ref. \cite{herrera_sant97}. Extended models, generalizing the results of \emph{N}-body simulations [see Eq.(\ref{n_body})] with the presence of constant anisotropy, can be found in Ref. \cite{ma_he08} and examples of theoretical models, described by the anisotropic version of Eq.(\ref{fd_cutoff}), are in Refs. \cite{1989ralston,1992imrs}). It is also important to mention models where the source of anisotropy is represented by the action of external fields \cite{canchiu_68a,canchiu_68b,2010fikps,2012dms}.

In this paper we study the gravitational equilibrium of static Newtonian configurations with spherical symmetry, composed by collisionless semidegenerate Fermi gas, in the presence of a cutoff term in the distribution function. We analyze different levels of anisotropies in the momentum phase space, addressing the relativistic solutions of anisotropic equilibrium models in a forthcoming paper. Finally, for the sake of completeness, it is important to stress the possibility of considering classical distribution functions, due to large values of the predicted mass of the particles \cite{deV_sanchez}. This question has been considered in the Appendix.

\section{Main Equations}

Let us consider the distribution function in the form
\begin{equation} \label{f_m_a}
\begin{cases}
f=\frac{g}{h^{3}} \Bigl(1 + \frac{L^{2}}{L_{c}^{2}}\Bigr)^l \frac{1 - e^{(\epsilon - \epsilon_{c})/kT}}{e^{(\epsilon -\mu)/kT}+1} & \text{for 
$\epsilon\leq\epsilon_{c}$} \\
f=0 & \text{for $\epsilon > \epsilon_{c}$}\ ,
\end{cases}
\end{equation}
where $L = mv_{t}r = p_{t}r$ is the angular momentum of the particle; $L_{c}= m\sigma r_{a}$ is a constant depending on the anisotropy radius $r_{a}$ (with $\sigma^{2} = 2kT/m$); $\epsilon_{c} = m(\Phi_{R} - \Phi)$ is the cutoff kinetic energy at a given radius $r$ from the center of the system ($\Phi$ is the gravitational potential undergone by a particle at distance $r$). This distribution function generalizes to a system of fermions the equilibrium solutions introduced by Bisnovatyi-Kogan, Merafina and Vaccarelli \cite{bkmv09,bkmv10}, hereafter named BKMV09 and BKMV10 respectively, for a system of nonquantum particles described by a Boltzmann distribution function with anisotropy and cutoff. In our case in the classical limit the term depending on energy recovers the well known King distribution function \cite{kingIII}
\begin{equation} \label{f_king}
\begin{cases}
f = \frac{g}{h^{3}}e^{\mu/kT}\Bigl(1 + \frac{L^{2}}{L_{c}^2}\Bigr)^l (e^{-\epsilon/kT}-e^{-\epsilon_{c}/kT})&\text{for $\epsilon\leq\epsilon_{c}$}\\
f = 0 & \text{for $\epsilon > \epsilon_{c}$}\ ,
\end{cases}
\end{equation}
In fact the King distribution function can be easily obtained from Eq.(\ref{f_king}) in the isotropic limit $L_c\rightarrow\infty$. The characteristic of models following the distribution function of Eq.(\ref{f_king}) will be studied in the present paper in the limit of $\mu/kT \rightarrow \infty$. Following Merafina and Ruffini \cite{meruf89,meruf90} let us introduce the variables
\begin{equation} \label{variables}
x = \frac{\epsilon}{kT}, \quad W = \frac{\epsilon_{c}}{kT}, \quad \theta = \frac{\mu}{kT}, \quad p_{r} = p \cos \alpha, \quad p_{t} = p \sin \alpha, \quad 0 \leq \alpha \leq \pi\ ,
\end{equation}
where $p_{r}$ and $p_{t}$ are, respectively, the radial and tangential component of the momentum of the particle. The relation between $W$ and the degeneracy parameter $\theta$ is given by
\begin{equation} \label{w_thetas}
W = \theta - \theta_{R}\ ,
\end{equation}
$W=0$ being at the edge of the equilibrium configuration, where $r=R$. The quantity $\theta_{R}$ denotes the value of the degeneracy parameter at the surface of the configuration. From the condition $\mu\leq\epsilon_c$ we obtain $\theta \leq W$ and $\theta_R \leq 0$ (in the fully degenerate case 
$\theta=W$ and $\theta_R=0$). Computing Eq.(\ref{w_thetas}) at the center of the configuration we obtain $\theta_{R} =\theta_0 -W_0$. Then thermodynamic quantities, describing ensemble of fermions, can be written as (see BKMV09)
\begin{equation} \label{n_state}
n = 2\pi \int fp_{t}dp_{t}dp_{r} =\frac{\pi gm^{3}\sigma^{3}}{h^{3}} \sum_{k=0}^{l} \binom{l}{k} \Bigl(\frac{r}{r_{a}}\Bigr)^{2k} A_{k} \int_0^W x^{k+\frac{1}{2}} g(x,W)dx\ ,
\end{equation}
\begin{equation} \label{radial_p}
P_{rr} = 2\pi \int f\frac{p_{r}^{2}}{m} p_{t}dp_{t}dp_{r} =\frac{\pi gm^{4}\sigma^{5}}{h^{3}} \sum_{k=0}^{l} \binom{l}{k} \Bigl(\frac{r}{r_{a}}\Bigr)^{2k}(A_{k}-A_{k+1}) \int_0^W x^{k+\frac{3}{2}} g(x,W)dx\ ,
\end{equation}
\begin{equation} \label{tangential_p}
P_{t} = \pi \int f\frac{p_{t}^{3}}{m}dp_{t}dp_{r} =\frac{\pi gm^{4}\sigma^{5}}{2h^{3}} \sum_{k=0}^{l} \binom{l}{k} \Bigl(\frac{r}{r_{a}}\Bigr)^{2k}A_{k+1} \int_0^W x^{k+\frac{3}{2}} g(x,W)dx\ ,
\end{equation}
where also $\rho = mn$ and the kinetic energy density \emph{u} is defined by $u =\frac{1}{2}P_{rr}+P_{t}$. Here the $A_{k}$ coefficients (\cite{1985bs_book}) and the $g(x,W)$ function are given, respectively, by
\begin{equation} \label{ak}
A_{k} = \int_0^\pi (\sin\alpha)^{2k+1} d\alpha = 2 \sum_{i=0}^{k} \binom{k}{i} \frac{(-1)^i}{2i+1}\ ,
\end{equation}
\begin{equation} \label{g_x_W}
g(x,W) = \frac{1 - e^{x-W}}{e^{x-\theta} + 1} = \frac{1 - e^{x-W}}{e^{x-W-\theta_{R}} + 1}\ .
\end{equation}
In particular, the first three values of $A_{k}$ coefficients $A_{0}$, $A_{1}$, and $A_{2}$ are 2, 4/3. and 16/15 respectively. Moreover, computing the preceding quantities, we have considered the Newtonian binomial relation
\begin{equation} \label{newton}
\Bigl(1+\frac{L^{2}}{L_{c}^{2}}\Bigr)^{l}=\sum_{k=0}^{l}\binom{l}{k} \Bigl(\frac{L}{L_{c}}\Bigr)^{2k}\ ,\quad\quad\mbox{with} \quad\binom{l}{k}= \frac{l!}{k!(l!-k!)}\ , \quad 0! = 1\ .
\end{equation}
Equilibrium equations for an anisotropic system are written in the form \cite{bkz69a}
\begin{equation} \label{eq_bkz}
\begin{cases}
\frac{dP_{rr}}{dr} & = -\frac{GM_{r}\rho}{r^{2}} - \frac{2}{r} (P_{rr} - P_{t}) \\
\frac{dM_{r}}{dr} & = 4\pi \rho(r) r^{2} \ ,
\end{cases}
\end{equation}
with conditions $P_{rr}(0) = P_{rr0}$ and $M_{r}(0) = 0$. The absence of an equation for $P_{t}$ arises from the isotropy of $P_t$ in the plane perpendicular to the direction of $P_{rr}$ and to the gravitational force. In this plane, we have an automatic compensation of the effects of the tangential pressure $P_t$ in every direction. Then, the equilibrium can be obtained by using only one equation involving $W$, defined in Eq.(\ref{variables}):
\begin{equation} \label{poisson_w}
\frac{d^2W}{dr^2} + \frac{2}{r}\frac{dW}{dr} = -\frac{8\pi G}{\sigma^{2}}\rho\ ,
\end{equation}
with boundary conditions $W(0) = W_{0}$ and $W'(0) = 0$.

\section{Numerical Results}

Following BKMV09, let us introduce nondimensional variables
\begin{equation} \label{dimensionless}
r = \xi \tilde{r}, \quad r_{a} = \xi a, \quad n = \frac{\sigma^2 \tilde{n}}{Gm\xi^{2}}, \quad \rho = \frac{\sigma^2 \tilde{\rho}}{G\xi^{2}}, \quad P_{rr} = \frac{\sigma^4 \tilde{P}_{rr}}{G\xi^{2}},
\quad P_{t} = \frac{\sigma^4 \tilde{P}_{t}}{G\xi^{2}}, \quad M_{r} = \frac{\sigma^2 \xi \tilde{M}_{r}}{G}\ ,
\end{equation}
where $\xi = (h^3/gG\sigma m^{4})^{1/2}$ and {\itshape a} is the anisotropy parameter. The nondimensional thermodynamic quantities become
\begin{equation} \label{n_state_less}
\tilde{n} = \tilde{\rho} = \pi \sum_{k=0}^{l} \binom{l}{k} \Bigl(\frac{\tilde{r}}{a}\Bigr)^{2k} A_{k} \int_0^W x^{k+\frac{1}{2}} g(x,W)dx\ ,
\end{equation}
\begin{equation} \label{radial_p_less}
\tilde{P}_{rr} = \pi \sum_{k=0}^{l} \binom{l}{k} \Bigl(\frac{\tilde{r}}{a}\Bigr)^{2k}(A_{k} - A_{k+1}) \int_0^W x^{k+\frac{3}{2}} g(x,W)dx\ ,
\end{equation}
\begin{equation} \label{tangential_p_less}
\tilde{P}_{t} = \frac{\pi}{2} \sum_{k=0}^{l} \binom{l}{k} \Bigl(\frac{\tilde{r}}{a}\Bigr)^{2k} A_{k+1} \int_0^W x^{k+\frac{3}{2}} g(x,W)dx\ ,
\end{equation}
and the equilibrium equation may be rewritten as
\begin{equation} \label{poisson_w_less}
\frac{d^2W}{d\tilde{r}^2} + \frac{2}{\tilde{r}}\frac{dW}{d\tilde{r}} = -8\pi \tilde{\rho}\ .
\end{equation}

Now, if we consider {\itshape l} = 1 in Eq.(\ref{f_m_a}) and calculate the equilibrium configurations for different values of the anisotropy parameter (we have chosen 1, 0.5, $10^{-1}$, $10^{-3}$ and $10^{-5}$) we obtain solutions varying $W_{0}$ and $\theta_{0}$. The level of the anisotropy in distribution function depends on the value of {\itshape a}. This can be represented by the ratio of radial to tangential mean square velocities (BKMV09)
\begin{equation} \label{eta}
\eta = \frac{2\langle v_{r}^{2} \rangle}{\langle v_{t}^{2} \rangle} = \frac{2\frac{P_{rr}}{\rho}}{\frac{2P_{t}}{\rho}} = \frac{P_{rr}}{P_{t}} = \frac{\tilde{P}_{rr}}{\tilde{P}_{t}} = \frac{1 + \frac{2}{5}{(\frac{\tilde{r}}{a})}^{2} \frac{I_{5/2}(W)}{I_{3/2}(W)}}{1 + \frac{4}{5}{(\frac{\tilde{r}}{a})}^{2} \frac{I_{5/2}(W)}{I_{3/2}(W)}} \leq 1\ ,
\end{equation}
where $I_{3/2}(W)$ and $I_{5/2}(W)$ are defined by the function
\begin{equation} \label{integrali}
I_n(W) = \int_0^W x^n g(x,W)dx\ .
\end{equation}
In Figs.\,1$-$5 we have represented the quantity $\eta$ as a function of the relative radius $r/R$ for five values of anisotropy parameter {\itshape a} and for different values of $W_{0}$ and $\theta_{0}$.
We can analyze the behavior of $\eta$ in the center of the configuration where $r = 0$ and at the edge where $r = R$. In both cases we have $\eta \rightarrow 1$. In the first case the result is obtained by Eq.(\ref{eta}) while in the second one we have $I_{5/2}(W)/I_{3/2}(W) \rightarrow 0$ being $W = 0$ at $r = R$.
Each figure shows a maximum value of $\eta$ in correspondence of the center of the configuration ($\eta$=1), indicating that isotropy in distribution of velocities prevails. Anisotropy becomes important towards the periphery of the configuration involving a decrease of $\eta$ that implies the prevalence of tangential motion. The quantity $\eta$ reaches its minimum value (which may be up to 0.5) for small values of relative radius. Then $\eta$ begins to increase until reaching the maximum value $\eta$ = 1. The thickness of the external isotropic region is rapidly decreasing with decreasing of the anisotropy parameter {\itshape a} (corresponding to a high level of anisotropy). Generally, for a distribution function like Eq.(\ref{f_m_a}) (with $l \geq 1$), the minimum value of $\eta$ is (BKMV09)
\begin{equation} \label{eta_min}
\eta_{min} = 2 \Bigl(\frac{A_{l}}{A_{l+1}} - 1\Bigr) = \frac{1}{l+1} \ ,
\end{equation}
and, for {\itshape l} = 1, we have $\eta_{\text{min}}$ = 0.5. For small values of $W_{0}$ and $\theta_{0}$ we obtain the same results of the classical case of BKMV09. Furthermore, a similar kind of the ``distribution of the motion," due to the presence of a certain degree of anisotropy, was found by \cite{nav_2010,takli97}. In these cases, the characteristics of the motion are analyzed by defining the parameter
\begin{equation} \label{beta}
\beta = 1 - \frac{\langle v_{t}^{2} \rangle}{\langle 2v_{r}^{2} \rangle} = 1 - \frac{1}{\eta}\ ,
\end{equation}
that has the same behavior of $\eta$.

\section{Density Profiles of Equilibrium Configurations}

By integrating Eq.(\ref{poisson_w_less}), we can describe density profiles of various configurations. The presence of the anisotropy parameter affects the behavior of the density function: in the limit $a \rightarrow 0$ we can see a general increase of the density and of its maximum value. In Figs.\,6$-$10 we have represented the quantity $\rho/\rho_{0}$ as function of the dimensionless radial coordinate $r/\xi$ for five values of anisotropy parameter {\itshape a} and for different values of $W_{0}$ and $\theta_{0}$.


The following figures show the behaviors of density profiles due to variation of $W_{0}$ and $\theta_{0}$. For high values of $\theta_{0}$ (degenerate case), we record an increase of the maximum value of the density function decreasing $\theta_{0}$ (coherently with the decrease of the degree of degeneration): we also note a displacement of the maximum in the direction of the periphery of configuration. For smaller (semidegenerate case) and negative (classical case) values of $\theta_{0}$ the behavior of density profiles is the same as the ones in BKMV09.
The behavior of density profiles is in accordance with Ralston and Smith's work \cite{1991ralsmith} concerning the presence of anisotropy in galactic halos (suggesting that these halos are formed by semidegenerate fermions). The analysis of Ralston and Smith stresses, moreover, the existence of hollow equilibrium configurations: confirmation at these features is found also in BKMV09, BKMV10, and Ref. \cite{nguyen_ped13}. These considerations lead us to conclude that the existence of these configurations depends on the presence of anisotropy in velocities, rather than choice of distribution function.
It seems from the figures that the value of the relative density approaches to zero at the center of the equilibrium configurations (especially at large anisotropy): it is a scale effect due a large scale used in values of $\rho/\rho_{0}$. This also shows that the central density is several orders of magnitude smaller than the maximum value of density function $\rho(r)$. The results of the numerical integrations giving the main parameters of the  different equilibrium configurations are reported in Tabs.\,I$-$V.

\section{Limits on the Particle Mass}

Limits on the value of the mass of the particles composing galactic halos may be derived from phase space constraints and in particular from the distribution function adopted (see, for example, \citep{1991ralsmith} and references therein). In particular, Tremaine and Gunn \cite{1979tregunn} derived a first limit within the realm of the classical statistics, while Gao and Ruffini \cite{gaoruf} established a second one by the assumption that galactic halos are composed by degenerate fermions. Following this last hypothesis, we derive an expression in which the mass of the particle is related explicitly to the anisotropy parameter \emph{a} (for a similar treatment on the same subject see \cite{1992imrs}). Let us rewrite the definition of the occupation number \emph{n}
\begin{equation} \label{n_bis}
n = \frac{\pi g m^{3} \sigma^{3}}{h^{3}} \sum_{k=0}^{l} \binom{l}{k} \Bigl(\frac{r}{r_{a}}\Bigr)^{2k} A_{k} \int_0^W x^{k+\frac{1}{2}}g(x,W)dx\ .
\end{equation}
In the limit of full degeneracy ($\theta \rightarrow W$ with $W \rightarrow \infty$ and thus $\theta_{R} \rightarrow 0$), $g(x,W) \rightarrow 1$ and Eq.(\ref{n_bis}) takes the form
\begin{equation} \label{n_ter}
n \leq \frac{\pi g m^{3} \sigma^{3}}{h^{3}} \sum_{k=0}^{l} \binom{l}{k} \Bigl(\frac{r}{r_{a}}\Bigr)^{2k} A_{k} \int_0^W x^{k+\frac{1}{2}}dx = \frac{2\pi g m^{3} \sigma^{3}}{h^{3}} \sum_{k=0}^{l} \binom{l}{k} \Bigl(\frac{r}{r_{a}}\Bigr)^{2k} A_{k} \frac{W^{k+\frac{3}{2}}}{2k+3}\ .
\end{equation}
The ratio $r/r_{a}$ can be expressed by the corresponding one in terms of the nondimensional variables ($\tilde{r}/a$) and the occupation number by the density (which is related by $\rho = mn$). Making the substitution, we have
\begin{equation} \label{n_rho}
\rho \leq \frac{2 \pi g m^{4} \sigma^{3}}{h^{3}} \sum_{k=0}^{l} \binom{l}{k} \Bigl(\frac{\tilde{r}}{a}\Bigr)^{2k} A_{k} \frac{W^{k+\frac{3}{2}}}{2k+3}
\end{equation}
and, for \emph{l} = 1, we obtain
\begin{equation} \label{rho_opetativa}
\rho \leq \frac{2 \pi g m^{4} \sigma^{3}}{h^{3}} \Bigl\{A_{0} \frac{W^{\frac{3}{2}}}{3} + A_{1} \Bigl(\frac{\tilde{r}}{a}\Bigr)^{2} \frac{W^{\frac{5}{2}}}{5}\Bigr\} = \frac{4\pi gm^{4}\sigma^{3} W^{\frac{3}{2}}}{3h^{3}} \Bigl\{1 + \frac{2}{5}\Bigl(\frac{\tilde{r}}{a}\Bigr)^{2}W\Bigr\}\ .
\end{equation}
Solving for the mass, we have therefore
\begin{equation} \label{massa_1}
m \geq \Biggl[\frac{3\rho h^{3}}{4\pi g\sigma^{3} W^{\frac{3}{2}}} \quad \frac{1}{1+\frac{2}{5}(\frac{\tilde{r}}{a})^{2}W}\Biggr]^{\frac{1}{4}}\ .
\end{equation}
By computing at the center of the ensemble,
\begin{equation} \label{massa_2}
m^{4} \geq \frac{3\rho_{0} h^{3}}{4\pi g\sigma^{3} W_{0}^{\frac{3}{2}}} \quad \quad \quad \rightarrow \quad \quad \quad m \geq \Biggl[\frac{3\rho_{0} h^{3}}{4\pi g\sigma^{3} W_{0}^{\frac{3}{2}}}\Biggr]^{\frac{1}{4}}\ .
\end{equation}
Assuming, for example, $\rho_{0} \sim 10^{-30}$ g/cm$^{3}$, which is the same order of magnitude of the critical density of the Universe (assuming the Hubble constant $H_{0} = 70$ km/s/Mpc), $\sigma \sim 100$ km/s, $g = 2$ (spin 1/2 particles) and $W_{0} = 20$, we obtain a lower limit for the mass of the particles $m \geq 2.18$ $\times$ $10^{-33}$ g $\approx 1.22$ eV. Unfortunately, this last evaluation does not take into account the effects of anisotropy into the mass. To do this, we may compute Eq.(\ref{massa_2}) at the core radius $r = r_{c}$ (or, equivalently, at $\tilde{r} = \tilde{r}_{c}$), defined as the radius at which the surface density (projected) corresponds to the half value of the surface central density, $\Sigma(r_{c}) = \Sigma_{0}/2$. Referring to $W_{c}$ as the value of $W$ at the core radius $r_c$, we have
\begin{equation} \label{massa_3}
m \geq \Biggl[\frac{3\rho(r_{c}) h^{3}}{4\pi g\sigma^{3} W_{c}^{\frac{3}{2}}} \quad \frac{1}{1 + \frac{2}{5}(\frac{r_{c}}{r_{a}})^{2}W_{c}}\Biggr]^{\frac{1}{4}}\ .
\end{equation}

Tabs.\,VI$-$x summarize the results obtained by applying Eq.(\ref{massa_3}) for different values of \emph{a}, $W_{0}$, $\rho(r_{c})$ and $\sigma$ and we note a double dependence of the values of \emph{m} by the anisotropy parameter and the central value of the cutoff energy. Starting from the anisotropic limit ($a = 10^{-5}$) towards the isotropic one ($a \geq 0.5$), we observe an increase of \emph{m} when the system recovers the isotropy in distribution of velocities. For any fixed value of \emph{a}, the lower limit of mass tends to decrease for increasing values of $W_{0}$ and this indicates that a further increase of the degeneracy level requires particles of small masses.


\section{Conclusions}

We have constructed models of anisotropic Newtonian ensembles of collisionless semidegenerate fermions with a distribution function which generalizes the Fermi-Dirac distribution with the energy cutoff, to describe the density profiles of equilibrium configurations which may be serious candidates for describing galactic dark halos. The distribution of dark matter in galactic halos may be estimated by using a method introduced by Persic and Salucci \cite{ps88}, i.e., the mass decomposition from rotational curves of galaxies. In fact, there is a clear connection between the disk-to-halo mass ratio, $M_{disk}/M_{halo}$, (at optical radii) and the luminosity $L_{B}$ being $M_{disk}/M_{halo} \propto L_{B}^{2/3}$ \cite{ps90a,ps90b}. This method therefore relates photometric measurements to distribution of matter.

The results displayed in Tabs.\,I$-$V and in the figures indicate clearly that, by varying values of \emph{a}, the anisotropy in the distribution of velocities of particles influences in a different way the behavior of the equilibrium configurations constructed. For high values of \emph{a} (1, 0.5), we note that the motion of particles is not influenced by anisotropy in a considerable way. The $\eta$ parameter unlikely reaches its minimum value even if, for \emph{a} = 0.5, in the intermediate zones of configurations ($0.4<r/R<0.8$), the descent arrives also to values 0.75 $-$ 0.8 (this is due to a decrease of the level of degeneracy within the ensemble). We can further note that the two values 1 and 0.5 do not affect in a considerable way the spatial extension of the particles and the total mass in the whole. From the tables we note indeed that the difference in terms of radii is very small (the discard is lower than 4$\%$) and still lower in the case of mass (we often note some values are practically the same). About density profiles, we recover the result of Ruffini and Stella \cite{1983rufs}. In the classical limit ($\theta_{0} \ll 0$) they tend to the typical behavior of the isothermal sphere and indeed the sizes of the ensembles of particles increase in a considerable way.

The situation becomes more interesting when \emph{a} = $10^{-1}$ because the anisotropy is more evident. From the figures we may note that the $\eta$ parameter approaches more its minimum value and for a larger spatial extension. The density profiles change drastically form, by evidencing the presence of hollow regions \cite{1991ralsmith,nguyen_ped13} and indicating that \emph{a} = $10^{-1}$ can make up a critical value for the trigger or not of ``hollowness." The results of the numerical integration show also the typical sizes and the mass of the configurations are smaller than the isotropic case, and a higher degree of anisotropy in the momentum distribution tends to ``condense" spatially the particles.

By still decreasing the value of the anisotropy parameter (\emph{a} = $10^{-3}$) we note as the hollowness of the density profiles is decidedly more evident. The maximum of the function $\rho/\rho_{0}$ increases in absolute value and its position is projected towards the peripheric zones of the ensembles of particles at decreasing the level of degeneracy. The motion of particles suffers more of the presence of the anisotropy. The two isotropic regions (central and peripheric) tend to become thinner and the decrease of $\eta$ from the maximum value to minimum is sharper than the preceding case. From data recorded in the tables we may see that a decrease by a factor of 100 of the anisotropy parameter implies a reduction of the radius and the mass by a factor about of 10, justifying the general increase of the density of the system and, in particular, of its maximum value. When \emph{a} = $10^{-5}$, the behavior of the density profiles and the $\eta$ parameter is more accentuated than the preceding case: the motion of particles becomes totally anisotropic and the density of the system still increases.

In the limit of full degeneracy (see Sec. 5), we have derived an expression in which the dependence of the mass by the anisotropy in distribution of momentum is explicit. In the limit $a \rightarrow 0$ and for values of matter density comprised between ($\sim 10^{-29} - \sim 10^{-17}$) g/cm$^{3}$ (with $W_{0} = 50$) we obtain $m \geq 0.05$ eV and $m \geq 48.6$ eV while, in the limit of complete isotropy ($a \geq 0.5$), for the same values of $\rho$ and $W_{0}$ we obtain $m \geq 0.6$ eV and $m \geq 616$ eV. The first three values of lower limit agree with the hypothesis that dark matter (both for galactic halos and for cosmic structures) is made by the lightest particles, like gravitinos and neutrinos, while the last value indicated is close to values proposed for more massive candidates for dark matter (e.g. axinos).

The level of degeneracy affects, as well as the degree of anisotropy, the behavior of the density profiles and we may summarize this as follows: from the degenerate systems (large values of $W_{0}$ and $\theta_{0}$) to semidegenerate ones ($\theta_{0} \sim 0$), we observe a shifting of the maximum of the density function towards the external regions of the equilibrium configuration, by noting an increase of the value of the maximum. Finally, for large and negative values of $\theta_{0}$ (classical limit), we recover the same result obtained by BKMV09.

\appendix
\section{If the particle mass is about a few keV; classical limit of quantum statistics?}

Cosmological models \cite{deV_sanchez} predict dark matter particles with mass $m_{*}$ in the keV scale. In this range, it should be possible to use the Boltzmann distribution function instead of the Fermi-Dirac one. For this reason, in this Appendix we will deduce an upper limit on the value of the degeneracy parameter at the center (i.e. $\theta_{0}$), by fixing the value of the mass of the particle, to verify if there exists the serious possibility to use the classical limit of the quantum statistics. In order to obtain an upper limit on $\theta_{0}$, let us consider the definition of the matter density, for both fermionic and classical particles, by using Eqs.(\ref{f_m_a}) and (\ref{f_king}), respectively, for $l=1$:
\begin{equation} \label{densities_1}
\begin{split}
\rho & = \frac{2\pi g m^{4} \sigma^{3}}{h^{3}}\Bigl[I_{1/2}(W)+\frac{2}{3} \Bigl(\frac{r}{r_{a}}\Bigr)^{2} I_{3/2}(W)\Bigr] \\
\rho & = \frac{4\pi g m^{4} \sigma^{3}}{3h^{3}}\,e^{\theta_R +W}\Bigl[\Gamma\Bigl(\frac{5}{2},W\Bigr)+\frac{2}{5}\Bigl(\frac{r}{r_{a}}\Bigr)^{2}\Gamma \Bigl(\frac{7}{2},W\Bigr)\Bigr]\ ,
\end{split}
\end{equation}
where $\Gamma(\frac{5}{2},W)$ and $\Gamma(\frac{7}{2},W)$ indicate the incomplete $\Gamma$ function
\begin{equation} \label{gamma_euler}
\Gamma(n,W) = \int_0^W e^{-x}x^{n-1} dx\ .
\end{equation}
By computing, for simplicity, Eqs.(\ref{densities_1}) at the center of the equilibrium configuration ($r=0$)
\begin{equation} \label{densities_2}
\begin{split}
\rho_{0} & = \frac{2\pi g m^{4} \sigma^{3}}{h^{3}}\,I_{1/2}(W_0) \\
\rho_{0} & = \frac{4\pi g m^{4}\sigma^{3}}{3h^{3}}\,e^{\theta_R +W_0}\,\Gamma\Bigl(\frac{5}{2},W_0\Bigr)\ .
\end{split}
\end{equation}
Finally, by using $\theta_{R} = \theta_{0} - W_{0}$ and isolating the mass of the particle,
\begin{equation} \label{densities_3}
\begin{split}
m^{4} & = \frac{\rho_{0}h^{3}}{2\pi g\sigma^{3}}\Bigl[I_{1/2}(W_0)\Bigr]^{-1}\\
m^{4} & = \frac{3\rho_{0}h^{3}}{4\pi g\sigma^{3}}\Bigl[e^{\theta_R+W_0}\,
\Gamma\Bigl(\frac{5}{2},W_0\Bigr)\Bigr]^{-1}\ .
\end{split}
\end{equation}
Since $m \leq m_{*}$, substituting in Eq.(\ref{densities_3}) we have
\begin{equation} \label{theta_0}
\begin{split}
& I_{1/2}(W_0)=\int_0^{W_0}\frac{1-e^{x-W_0}}{e^{x-\theta_0}+1}\,x^{1/2}dx
\geq \frac{\rho_{0} h^{3}}{2\pi g m_{*}^{4} \sigma^{3}} \\
& \theta_{0} \geq \ln \Bigl(\frac{3\rho_{0} h^{3}}{4\pi g m_{*}^{4} \sigma^{3}}\Bigr) - \ln \Bigl[\Gamma\Bigl(\frac{5}{2},W_0\Bigr)\Bigr]\ .
\end{split}
\end{equation}
The first inequality can be solved at fixed values of $W_0$ through the calculation of the integral which becomes a trivial function of $\theta_0$; conversely, the second one, referring to the classical limit, is directly solvable. In Table xI, we calculate the inequalities (\ref{theta_0}) for different values of the parameters. There are no appreciable differences between the quantum and the classical treatment regarding the values of the degeneracy parameter at the center of the equilibrium configurations. This leads us to conclude that the two points of view can be considered equivalent with respect to the analysis connected to the prediction of the mass of the particles composing the halo.

\bibliography{biblio}

\clearpage

\begin{longtable}{>{\centering\arraybackslash}p{2.5cm}>{\centering\arraybackslash}p{2.5cm}>{\centering\arraybackslash}p{2.5cm}>{\centering\arraybackslash}p{2.5cm}>{\centering\arraybackslash}p{2.5cm}}
\caption{Numerical characteristics for semidegenerate fermions for {\itshape a} = 1 and for different values of $W_{0}$ and $\theta_{0}$. $\tilde{R}$ and $\tilde{M}$ are, respectively, dimensionless radius and mass of equilibrium configurations.}
\label{tab:longtable1} \\
\hline
\hline
$W_{0}$ & $\theta_{0}$ & $\theta_{R}$ & $\tilde{R}$ & $\tilde{M}$\\ 
\hline
\endfirsthead
\multicolumn{5}{c}{Table I (\emph{continued}).} \\
\hline
\hline
$W_{0}$ & $\theta_{0}$ & $\theta_{R}$ & $\tilde{R}$ & $\tilde{M}$ \\ 
\hline
\endhead
\hline
\endfoot
\hline
\endlastfoot
1 & 1 & 0 & 1.15 $\times$ $10^{0}$ & 2.42 $\times$ $10^{-1}$ \\ 
 & 0 & $-$1 & 1.51 $\times$ $10^{0}$ & 3.05 $\times$ $10^{-1}$ \\ 
 & $-$1 & $-$2 & 2.16 $\times$ $10^{0}$ & 4.33 $\times$ $10^{-1}$ \\
3 & 3 & 0 & 5.17 $\times$ $10^{-1}$ & 3.51 $\times$ $10^{-1}$ \\  
 & 1 & $-$2 & 9.01 $\times$ $10^{-1}$ & 4.78 $\times$ $10^{-1}$ \\  
 & 0 & $-$3 & 1.36 $\times$ $10^{0}$ & 6.57 $\times$ $10^{-1}$ \\ 
5 & 5 & 0 & 3.68 $\times$ $10^{-1}$ & 4.55 $\times$ $10^{-1}$ \\  
 & 3 & $-$2 & 5.71 $\times$ $10^{-1}$ & 5.15 $\times$ $10^{-1}$ \\  
 & 1 & $-$4 & 1.38 $\times$ $10^{0}$ & 8.11 $\times$ $10^{-1}$ \\  
 & 0 & $-$5 & 2.18 $\times$ $10^{0}$ & 1.20 $\times$ $10^{0}$ \\ 
7 & 7 & 0 & 3.01 $\times$ $10^{-1}$ & 5.61 $\times$ $10^{-1}$ \\  
 & 5 & $-$2 & 4.20 $\times$ $10^{-1}$ & 5.82 $\times$ $10^{-1}$ \\  
 & 3 & $-$4 & 9.48 $\times$ $10^{-1}$ & 7.13 $\times$ $10^{-1}$ \\  
 & 1 & $-$6 & 2.96 $\times$ $10^{0}$ & 1.37 $\times$ $10^{0}$ \\ 
 & 0 & $-$7 & 4.22 $\times$ $10^{0}$ & 2.15 $\times$ $10^{0}$ \\ 
10 & 10 & 0 & 2.50 $\times$ $10^{-1}$ & 7.18 $\times$ $10^{-1}$ \\  
 & 7 & $-$3 & 3.95 $\times$ $10^{-1}$ & 7.41 $\times$ $10^{-1}$ \\  
 & 5 & $-$5 & 9.54 $\times$ $10^{-1}$ & 7.70 $\times$ $10^{-1}$ \\
 & 3 & $-$7 & 5.72 $\times$ $10^{0}$ & 1.46 $\times$ $10^{0}$ \\
 & 1 & $-$9 & 8.56 $\times$ $10^{0}$ & 4.12 $\times$ $10^{0}$ \\ 
 & 0 & $-$10 & 9.72 $\times$ $10^{0}$ & 5.71 $\times$ $10^{0}$ \\
15 & 15 & 0 & 2.09 $\times$ $10^{-1}$ & 9.68 $\times$ $10^{-1}$ \\  
 & 10 & $-$5 & 4.33 $\times$ $10^{-1}$ & 9.92 $\times$ $10^{-1}$ \\ 
20 & 20 & 0 & 1.86 $\times$ $10^{-1}$ & 1.20 $\times$ $10^{0}$ \\  
 & 15 & $-$5 & 2.98 $\times$ $10^{-1}$ & 1.39 $\times$ $10^{0}$ \\ 
30 & 30 & 0 & 1.61 $\times$ $10^{-1}$ & 1.64 $\times$ $10^{0}$ \\  
 & 20 & $-$10 & 4.40 $\times$ $10^{-1}$ & 1.92 $\times$ $10^{0}$ \\ 
50 & 50 & 0 & 1.37 $\times$ $10^{-1}$ & 2.42 $\times$ $10^{0}$ \\  
 & 30 & $-$20 & 1.07 $\times$ $10^{0}$ & 3.72 $\times$ $10^{0}$ \\ 
\end{longtable}

\begin{longtable}{>{\centering\arraybackslash}p{2.5cm}>{\centering\arraybackslash}p{2.5cm}>{\centering\arraybackslash}p{2.5cm}>{\centering\arraybackslash}p{2.5cm}>{\centering\arraybackslash}p{2.5cm}}
\caption{The same as Table I, for \emph{a} = 0.5.}
\label{tab:longtable2} \\
\hline
\hline
$W_{0}$ & $\theta_{0}$ & $\theta_{R}$ & $\tilde{R}$ & $\tilde{M}$\\ 
\hline
\endfirsthead
\multicolumn{5}{c}{Table II (\emph{continued}).} \\
\hline
\hline
$W_{0}$ & $\theta_{0}$ & $\theta_{R}$ & $\tilde{R}$ & $\tilde{M}$ \\ 
\hline
\endhead
\hline
\endfoot
\hline
\endlastfoot
1 & 1 & 0 & 1.09 $\times$ $10^{0}$ & 2.41 $\times$ $10^{-1}$ \\ 
 & 0 & $-$1 & 1.40 $\times$ $10^{0}$ & 3.03 $\times$ $10^{-1}$ \\ 
 & $-$1 & $-$2 & 1.90 $\times$ $10^{0}$ & 4.26 $\times$ $10^{-1}$ \\
 & $-$3 & $-$4 & 3.43 $\times$ $10^{0}$ & 9.48 $\times$ $10^{-1}$ \\ 
 & $-$5 & $-$6 & 5.70 $\times$ $10^{0}$ & 1.92 $\times$ $10^{0}$ \\ 
 & $-$7 & $-$8 & 9.34 $\times$ $10^{0}$ & 3.46 $\times$ $10^{0}$ \\
 & $-$10 & $-$11 & 1.97 $\times$ $10^{1}$ & 7.66 $\times$ $10^{0}$ \\  
3 & 3 & 0 & 5.00 $\times$ $10^{-1}$ & 3.50 $\times$ $10^{-1}$ \\  
 & 1 & $-$2 & 8.38 $\times$ $10^{-1}$ & 4.76 $\times$ $10^{-1}$ \\  
 & 0 & $-$3 & 1.20 $\times$ $10^{0}$ & 6.50 $\times$ $10^{-1}$ \\
 & $-$1 & $-$4 & 1.67 $\times$ $10^{0}$ & 9.57 $\times$ $10^{-1}$ \\
 & $-$3 & $-$6 & 2.83 $\times$ $10^{0}$ & 2.10 $\times$ $10^{0}$ \\
 & $-$5 & $-$8 & 4.56 $\times$ $10^{0}$ & 4.04 $\times$ $10^{0}$ \\
 & $-$7 & $-$10 & 7.40 $\times$ $10^{0}$ & 7.12 $\times$ $10^{0}$ \\
 & $-$10 & $-$13 & 1.56 $\times$ $10^{1}$ & 1.56 $\times$ $10^{1}$ \\ 
5 & 5 & 0 & 3.57 $\times$ $10^{-1}$ & 4.54 $\times$ $10^{-1}$ \\  
 & 3 & $-$2 & 5.42 $\times$ $10^{-1}$ & 5.14 $\times$ $10^{-1}$ \\  
 & 1 & $-$4 & 1.18 $\times$ $10^{0}$ & 8.04 $\times$ $10^{-1}$ \\  
 & 0 & $-$5 & 1.67 $\times$ $10^{0}$ & 1.16 $\times$ $10^{0}$ \\ 
7 & 7 & 0 & 2.92 $\times$ $10^{-1}$ & 5.59 $\times$ $10^{-1}$ \\
 & 5 & $-$2 & 4.03 $\times$ $10^{-1}$ & 5.80 $\times$ $10^{-1}$ \\
 & 3 & $-$4 & 8.47 $\times$ $10^{-1}$ & 7.12 $\times$ $10^{-1}$ \\  
 & 1 & $-$6 & 2.06 $\times$ $10^{0}$ & 1.31 $\times$ $10^{0}$ \\ 
 & 0 & $-$7 & 2.68 $\times$ $10^{0}$ & 1.92 $\times$ $10^{0}$ \\ 
10 & 10 & 0 & 2.42 $\times$ $10^{-1}$ & 7.14 $\times$ $10^{-1}$ \\  
 & 7 & $-$3 & 3.76 $\times$ $10^{-1}$ & 7.40 $\times$ $10^{-1}$ \\  
 & 5 & $-$5 & 8.44 $\times$ $10^{-1}$ & 7.70 $\times$ $10^{-1}$ \\  
 & 3 & $-$7 & 3.23 $\times$ $10^{0}$ & 1.31 $\times$ $10^{0}$ \\  
 & 1 & $-$9 & 5.13 $\times$ $10^{0}$ & 2.99 $\times$ $10^{0}$ \\  
 & 0 & $-$10 & 5.99 $\times$ $10^{0}$ & 4.02 $\times$ $10^{0}$ \\ 
15 & 15 & 0 & 2.01 $\times$ $10^{-1}$ & 9.62 $\times$ $10^{-1}$ \\  
 & 10 & $-$5 & 4.06 $\times$ $10^{-1}$ & 9.68 $\times$ $10^{-1}$ \\
 & 7 & $-$8 & 4.66 $\times$ $10^{0}$ & 9.72 $\times$ $10^{-1}$ \\
 & 5 & $-$10 & 8.38 $\times$ $10^{0}$ & 4.66 $\times$ $10^{0}$ \\
 & 3 & $-$12 & 1.02 $\times$ $10^{1}$ & 7.96 $\times$ $10^{0}$ \\
 & 1 & $-$14 & 1.61 $\times$ $10^{1}$ & 1.20 $\times$ $10^{1}$ \\
 & 0 & $-$15 & 2.13 $\times$ $10^{1}$ & 1.49 $\times$ $10^{1}$ \\
20 & 20 & 0 & 1.79 $\times$ $10^{-1}$ & 1.19 $\times$ $10^{0}$ \\  
 & 15 & $-$5 & 2.82 $\times$ $10^{-1}$ & 1.28 $\times$ $10^{0}$ \\ 
 & 10 & $-$10 & 5.98 $\times$ $10^{0}$ & 1.57 $\times$ $10^{0}$ \\
30 & 30 & 0 & 1.54 $\times$ $10^{-1}$ & 1.62 $\times$ $10^{0}$ \\  
 & 20 & $-$10 & 4.03 $\times$ $10^{-1}$ & 1.90 $\times$ $10^{0}$ \\ 
50 & 50 & 0 & 1.39 $\times$ $10^{-1}$ & 2.39 $\times$ $10^{0}$ \\  
 & 30 & $-$20 & 8.49 $\times$ $10^{-1}$ & 3.71 $\times$ $10^{0}$ \\ 
\end{longtable}

\begin{longtable}{>{\centering\arraybackslash}p{2.5cm}>{\centering\arraybackslash}p{2.5cm}>{\centering\arraybackslash}p{2.5cm}>{\centering\arraybackslash}p{2.5cm}>{\centering\arraybackslash}p{2.5cm}}
\caption{The same as Table I, for \emph{a} = $10^{-1}$.}
\label{tab:longtable3} \\
\hline
\hline
$W_{0}$ & $\theta_{0}$ & $\theta_{R}$ & $\tilde{R}$ & $\tilde{M}$\\ 
\hline
\endfirsthead
\multicolumn{5}{c}{Table III (\emph{continued}).} \\
\hline
\hline
$W_{0}$ & $\theta_{0}$ & $\theta_{R}$ & $\tilde{R}$ & $\tilde{M}$ \\ 
\hline
\endhead
\hline
\endfoot
\hline
\endlastfoot
1 & 1 & 0 & 6.76 $\times$ $10^{-1}$ & 2.08 $\times$ $10^{-1}$ \\ 
 & 0 & $-$1 & 7.26 $\times$ $10^{-1}$ & 2.23 $\times$ $10^{-1}$ \\ 
 & $-$1 & $-$2 & 9.70 $\times$ $10^{-1}$ & 3.19 $\times$ $10^{-1}$ \\
 & $-$3 & $-$4 & 1.55 $\times$ $10^{0}$ & 5.61 $\times$ $10^{-1}$ \\ 
 & $-$5 & $-$6 & 2.53 $\times$ $10^{0}$ & 9.64 $\times$ $10^{-1}$ \\ 
 & $-$7 & $-$8 & 4.16 $\times$ $10^{0}$ & 1.62 $\times$ $10^{0}$ \\
 & $-$10 & $-$11 & 8.80 $\times$ $10^{0}$ & 3.46 $\times$ $10^{0}$ \\  
3 & 3 & 0 & 3.35 $\times$ $10^{-1}$ & 3.13 $\times$ $10^{-1}$ \\  
 & 1 & $-$2 & 4.72 $\times$ $10^{-1}$ & 4.00 $\times$ $10^{-1}$ \\  
 & 0 & $-$3 & 5.96 $\times$ $10^{-1}$ & 5.03 $\times$ $10^{-1}$ \\
 & $-$1 & $-$4 & 7.57 $\times$ $10^{-1}$ & 6.60 $\times$ $10^{-1}$ \\
 & $-$3 & $-$6 & 1.22 $\times$ $10^{0}$ & 1.16 $\times$ $10^{0}$ \\
 & $-$5 & $-$8 & 2.00 $\times$ $10^{0}$ & 1.97 $\times$ $10^{0}$ \\
 & $-$7 & $-$10 & 3.28 $\times$ $10^{0}$ & 3.29 $\times$ $10^{0}$ \\
 & $-$10 & $-$13 & 6.94 $\times$ $10^{0}$ & 7.01 $\times$ $10^{0}$ \\ 
5 & 5 & 0 & 2.43 $\times$ $10^{-1}$ & 4.04 $\times$ $10^{-1}$ \\  
 & 3 & $-$2 & 3.27 $\times$ $10^{-1}$ & 4.48 $\times$ $10^{-1}$ \\  
 & 1 & $-$4 & 5.39 $\times$ $10^{-1}$ & 6.22 $\times$ $10^{-1}$ \\  
 & 0 & $-$5 & 6.95 $\times$ $10^{-1}$ & 7.99 $\times$ $10^{-1}$ \\
 & $-$1 & $-$6 & 8.84 $\times$ $10^{-1}$ & 1.05 $\times$ $10^{0}$ \\
 & $-$3 & $-$8 & 1.42 $\times$ $10^{0}$ & 1.81 $\times$ $10^{0}$ \\ 
 & $-$5 & $-$10 & 2.32 $\times$ $10^{0}$ & 3.04 $\times$ $10^{0}$ \\
 & $-$7 & $-$12 & 3.80 $\times$ $10^{0}$ & 5.05 $\times$ $10^{0}$ \\
 & $-$10 & $-$15 & 8.03 $\times$ $10^{0}$ & 1.08 $\times$ $10^{1}$ \\
7 & 7 & 0 & 1.99 $\times$ $10^{-1}$ & 4.93 $\times$ $10^{-1}$ \\  
 & 5 & $-$2 & 2.54 $\times$ $10^{-1}$ & 5.09 $\times$ $10^{-1}$ \\  
 & 3 & $-$4 & 4.13 $\times$ $10^{-1}$ & 5.94 $\times$ $10^{-1}$ \\  
 & 1 & $-$6 & 7.39 $\times$ $10^{-1}$ & 8.76 $\times$ $10^{-1}$ \\ 
 & 0 & $-$7 & 9.53 $\times$ $10^{-1}$ & 1.13 $\times$ $10^{0}$ \\
 & $-$1 & $-$8 & 1.21 $\times$ $10^{0}$ & 1.48 $\times$ $10^{0}$ \\
 & $-$3 & $-$10 & 1.93 $\times$ $10^{0}$ & 2.49 $\times$ $10^{0}$ \\
 & $-$5 & $-$12 & 3.13 $\times$ $10^{0}$ & 4.16 $\times$ $10^{0}$ \\
 & $-$7 & $-$14 & 5.13 $\times$ $10^{0}$ & 6.90 $\times$ $10^{0}$ \\
 & $-$10 & $-$17 & 1.08 $\times$ $10^{1}$ & 1.47 $\times$ $10^{1}$ \\
10 & 10 & 0 & 1.65 $\times$ $10^{-1}$ & 6.22 $\times$ $10^{-1}$ \\  
 & 7 & $-$3 & 2.29 $\times$ $10^{-1}$ & 6.37 $\times$ $10^{-1}$ \\  
 & 5 & $-$5 & 3.86 $\times$ $10^{-1}$ & 6.51 $\times$ $10^{-1}$ \\  
 & 3 & $-$7 & 8.29 $\times$ $10^{-1}$ & 8.34 $\times$ $10^{-1}$ \\  
 & 1 & $-$9 & 1.60 $\times$ $10^{0}$ & 1.35 $\times$ $10^{0}$ \\  
 & 0 & $-$10 & 2.06 $\times$ $10^{0}$ & 1.74 $\times$ $10^{0}$ \\
 & $-$1 & $-$11 & 2.59 $\times$ $10^{0}$ & 2.24 $\times$ $10^{0}$ \\
 & $-$3 & $-$13 & 4.13 $\times$ $10^{0}$ & 3.68 $\times$ $10^{0}$ \\
 & $-$5 & $-$15 & 6.68 $\times$ $10^{0}$ & 6.06 $\times$ $10^{0}$ \\
 & $-$7 & $-$17 & 1.09 $\times$ $10^{1}$ & 9.99 $\times$ $10^{0}$ \\
15 & 15 & 0 & 1.35 $\times$ $10^{-1}$ & 8.21 $\times$ $10^{-1}$ \\  
 & 10 & $-$5 & 2.29 $\times$ $10^{-1}$ & 8.66 $\times$ $10^{-1}$ \\
 & 7 & $-$8 & 6.67 $\times$ $10^{-1}$ & 9.43 $\times$ $10^{-1}$ \\
 & 5 & $-$10 & 3.95 $\times$ $10^{0}$ & 1.15 $\times$ $10^{0}$ \\
 & 3 & $-$12 & 5.78 $\times$ $10^{0}$ & 2.94 $\times$ $10^{0}$ \\
 & 1 & $-$14 & 8.62 $\times$ $10^{0}$ & 5.02 $\times$ $10^{0}$ \\
 & 0 & $-$15 & 1.11 $\times$ $10^{1}$ & 6.43 $\times$ $10^{0}$ \\
20 & 20 & 0 & 1.18 $\times$ $10^{-1}$ & 1.00 $\times$ $10^{0}$ \\  
 & 15 & $-$5 & 1.69 $\times$ $10^{-1}$ & 1.96 $\times$ $10^{0}$ \\
 & 10 & $-$10 & 7.36 $\times$ $10^{-1}$ & 3.93 $\times$ $10^{0}$ \\
 & 7 & $-$13 & 7.71 $\times$ $10^{0}$ & 5.00 $\times$ $10^{0}$ \\
 & 5 & $-$15 & 9.00 $\times$ $10^{0}$ & 8.06 $\times$ $10^{0}$ \\
 & 3 & $-$17 & 1.42 $\times$ $10^{1}$ & 1.19 $\times$ $10^{1}$ \\
 & 1 & $-$19 & 2.39 $\times$ $10^{1}$ & 1.89 $\times$ $10^{1}$ \\
 & 0 & $-$20 & 3.07 $\times$ $10^{1}$ & 2.43 $\times$ $10^{1}$ \\
30 & 30 & 0 & 9.92 $\times$ $10^{-2}$ & 1.33 $\times$ $10^{0}$ \\  
 & 20 & $-$10 & 2.04 $\times$ $10^{-1}$ & 2.09 $\times$ $10^{0}$ \\ 
 & 15 & $-$15 & 1.58 $\times$ $10^{0}$ & 2.32 $\times$ $10^{0}$ \\ 
50 & 50 & 0 & 8.06 $\times$ $10^{-2}$ & 1.90 $\times$ $10^{0}$ \\  
 & 30 & $-$20 & 2.70 $\times$ $10^{-1}$ & 2.40 $\times$ $10^{0}$ \\ 
\end{longtable}

\begin{longtable}{>{\centering\arraybackslash}p{2.5cm}>{\centering\arraybackslash}p{2.5cm}>{\centering\arraybackslash}p{2.5cm}>{\centering\arraybackslash}p{2.5cm}>{\centering\arraybackslash}p{2.5cm}}
\caption{The same as Table I, for \emph{a} = $10^{-3}$.}
\label{tab:longtable4} \\
\hline
\hline
$W_{0}$ & $\theta_{0}$ & $\theta_{R}$ & $\tilde{R}$ & $\tilde{M}$\\ 
\hline
\endfirsthead
\multicolumn{5}{c}{Table IV (\emph{continued}).} \\
\hline
\hline
$W_{0}$ & $\theta_{0}$ & $\theta_{R}$ & $\tilde{R}$ & $\tilde{M}$ \\ 
\hline
\endhead
\hline
\endfoot
\hline
\endlastfoot
1 & 1 & 0 & 6.87 $\times$ $10^{-2}$ & 2.84 $\times$ $10^{-2}$ \\ 
 & 0 & $-$1 & 7.32 $\times$ $10^{-2}$ & 2.99 $\times$ $10^{-2}$ \\ 
 & $-$1 & $-$2 & 9.63 $\times$ $10^{-2}$ & 3.83 $\times$ $10^{-2}$ \\
 & $-$3 & -4 & 1.54 $\times$ $10^{-1}$ & 6.04 $\times$ $10^{-2}$ \\ 
 & $-$5 & $-$6 & 2.52 $\times$ $10^{-1}$ & 9.91 $\times$ $10^{-2}$ \\ 
 & $-$7 & -8 & 4.15 $\times$ $10^{-1}$ & 1.64 $\times$ $10^{-1}$ \\
 & $-$10 & $-$11 & 8.79 $\times$ $10^{-1}$ & 3.47 $\times$ $10^{-1}$ \\
 & $-$15 & $-$16 & 3.07 $\times$ $10^{0}$ & 1.21 $\times$ $10^{0}$ \\
3 & 3 & 0 & 3.48 $\times$ $10^{-2}$ & 4.43 $\times$ $10^{-2}$ \\  
 & 1 & -2 & 4.66 $\times$ $10^{-2}$ & 5.21 $\times$ $10^{-2}$ \\  
 & 0 & $-$3 & 5.79 $\times$ $10^{-2}$ & 6.11 $\times$ $10^{-2}$ \\
 & $-$1 & $-$4 & 7.35 $\times$ $10^{-2}$ & 7.54 $\times$ $10^{-2}$ \\
 & $-$3 & $-$6 & 1.21 $\times$ $10^{-1}$ & 1.22 $\times$ $10^{-1}$ \\
 & $-$5 & $-$8 & 1.99 $\times$ $10^{-1}$ & 2.01 $\times$ $10^{-1}$ \\
 & $-$7 & $-$10 & 3.28 $\times$ $10^{-1}$ & 3.31 $\times$ $10^{-1}$ \\
 & $-$10 & $-$13 & 6.93 $\times$ $10^{-1}$ & 7.02 $\times$ $10^{-1}$ \\
 & $-$15 & $-$18 & 2.42 $\times$ $10^{0}$ & 2.45 $\times$ $10^{0}$ \\
 & $-$20 & $-$23 & 8.45 $\times$ $10^{0}$ & 8.56 $\times$ $10^{0}$ \\
5 & 5 & 0 & 2.57 $\times$ $10^{-2}$ & 5.67 $\times$ $10^{-2}$ \\  
 & 3 & $-$2 & 3.26 $\times$ $10^{-2}$ & 6.05 $\times$ $10^{-2}$ \\  
 & 1 & $-$4 & 5.06 $\times$ $10^{-2}$ & 7.57 $\times$ $10^{-2}$ \\  
 & 0 & $-$5 & 6.51 $\times$ $10^{-2}$ & 9.10 $\times$ $10^{-2}$ \\
 & $-$1 & $-$6 & 8.42 $\times$ $10^{-2}$ & 1.14 $\times$ $10^{-1}$ \\
 & $-$3 & $-$8 & 1.39 $\times$ $10^{-1}$ & 1.86 $\times$ $10^{-1}$ \\
 & $-$5 & $-$10 & 2.30 $\times$ $10^{-1}$ & 3.07 $\times$ $10^{-1}$ \\
 & $-$7 & $-$12 & 3.79 $\times$ $10^{-1}$ & 5.07 $\times$ $10^{-1}$ \\
 & $-$10 & $-$15 & 8.02 $\times$ $10^{-1}$ & 1.08 $\times$ $10^{0}$ \\
 & $-$15 & $-$20 & 2.80 $\times$ $10^{0}$ & 3.76 $\times$ $10^{0}$ \\
 & $-$20 & $-$25 & 9.77 $\times$ $10^{0}$ & 1.31 $\times$ $10^{1}$ \\
7 & 7 & 0 & 2.12 $\times$ $10^{-2}$ & 6.79 $\times$ $10^{-2}$ \\  
 & 5 & $-$2 & 2.57 $\times$ $10^{-2}$ & 6.92 $\times$ $10^{-2}$ \\  
 & 3 & $-$4 & 3.82 $\times$ $10^{-2}$ & 7.60 $\times$ $10^{-2}$ \\  
 & 1 & $-$6 & 6.58 $\times$ $10^{-2}$ & 9.97 $\times$ $10^{-2}$ \\ 
 & 0 & $-$7 & 8.64 $\times$ $10^{-1}$ & 1.22 $\times$ $10^{-1}$ \\
 & $-$1 & $-$8 & 1.13 $\times$ $10^{-1}$ & 1.55 $\times$ $10^{-1}$ \\
 & $-$3 & $-$10 & 1.88 $\times$ $10^{-1}$ & 2.53 $\times$ $10^{-1}$ \\
 & $-$5 & $-$12 & 3.10 $\times$ $10^{-1}$ & 4.18 $\times$ $10^{-1}$ \\
 & $-$7 & $-$14 & 5.11 $\times$ $10^{-1}$ & 6.90 $\times$ $10^{-1}$ \\
 & $-$10 & $-$17 & 1.08 $\times$ $10^{0}$ & 1.47 $\times$ $10^{0}$ \\
 & $-$15 & $-$22 & 3.78 $\times$ $10^{0}$ & 5.11 $\times$ $10^{0}$ \\
 & $-$20 & $-$27 & 1.32 $\times$ $10^{1}$ & 1.79 $\times$ $10^{1}$ \\
10 & 10 & 0 & 1.76 $\times$ $10^{-2}$ & 8.37 $\times$ $10^{-2}$ \\  
 & 7 & $-$3 & 2.29 $\times$ $10^{-2}$ & 8.39 $\times$ $10^{-2}$ \\  
 & 5 & $-$5 & 3.44 $\times$ $10^{-2}$ & 8.40 $\times$ $10^{-2}$ \\  
 & 3 & $-$7 & 6.44 $\times$ $10^{-2}$ & 9.62 $\times$ $10^{-2}$ \\  
 & 1 & $-$9 & 1.32 $\times$ $10^{-1}$ & 1.37 $\times$ $10^{-1}$ \\  
 & 0 & $-$10 & 1.80 $\times$ $10^{-1}$ & 1.73 $\times$ $10^{-1}$ \\
 & $-$1 & $-$11 & 2.38 $\times$ $10^{-1}$ & 2.21 $\times$ $10^{-1}$ \\
 & $-$3 & $-$13 & 3.99 $\times$ $10^{-1}$ & 3.66 $\times$ $10^{-1}$ \\
 & $-$5 & $-$15 & 6.59 $\times$ $10^{-1}$ & 6.04 $\times$ $10^{-1}$ \\
 & $-$7 & $-$17 & 1.09 $\times$ $10^{0}$ & 9.97 $\times$ $10^{-1}$ \\
 & $-$10 & $-$20 & 2.30 $\times$ $10^{0}$ & 2.11 $\times$ $10^{0}$ \\
 & $-$15 & $-$25 & 8.03 $\times$ $10^{0}$ & 7.38 $\times$ $10^{0}$ \\
15 & 15 & 0 & 1.44 $\times$ $10^{-2}$ & 1.07 $\times$ $10^{-1}$ \\  
 & 10 & $-$5 & 2.21 $\times$ $10^{-2}$ & 1.38 $\times$ $10^{-1}$ \\
 & 7 & $-$8 & 4.57 $\times$ $10^{-2}$ & 1.87 $\times$ $10^{-1}$ \\
 & 5 & $-$10 & 1.44 $\times$ $10^{-1}$ & 1.98 $\times$ $10^{-1}$ \\
 & 3 & $-$12 & 7.29 $\times$ $10^{-1}$ & 2.26 $\times$ $10^{-1}$ \\
 & 1 & $-$14 & 9.73 $\times$ $10^{-1}$ & 4.84 $\times$ $10^{-1}$ \\
 & 0 & $-$15 & 1.20 $\times$ $10^{0}$ & 6.36 $\times$ $10^{-1}$ \\
 & $-$1 & $-$16 & 1.51 $\times$ $10^{0}$ & 8.22 $\times$ $10^{-1}$ \\
 & $-$3 & $-$18 & 2.47 $\times$ $10^{0}$ & 1.36 $\times$ $10^{0}$ \\
 & $-$5 & $-$20 & 4.06 $\times$ $10^{0}$ & 2.24 $\times$ $10^{0}$ \\
 & $-$7 & $-$22 & 6.70 $\times$ $10^{0}$ & 3.70 $\times$ $10^{0}$ \\
 & $-$10 & $-$25 & 1.42 $\times$ $10^{1}$ & 7.84 $\times$ $10^{0}$ \\
20 & 20 & 0 & 1.26 $\times$ $10^{-2}$ & 1.27 $\times$ $10^{-1}$ \\  
 & 15 & $-$5 & 1.69 $\times$ $10^{-2}$ & 1.57 $\times$ $10^{-1}$ \\
 & 10 & $-$10 & 4.37 $\times$ $10^{-2}$ & 2.02 $\times$ $10^{-1}$ \\
 & 7 & $-$13 & 1.86 $\times$ $10^{0}$ & 2.65 $\times$ $10^{-1}$ \\
 & 5 & $-$15 & 2.06 $\times$ $10^{0}$ & 8.49 $\times$ $10^{-1}$ \\
 & 3 & $-$17 & 2.13 $\times$ $10^{0}$ & 1.30 $\times$ $10^{0}$ \\
 & 1 & $-$19 & 2.32 $\times$ $10^{0}$ & 1.96 $\times$ $10^{0}$ \\
 & 0 & $-$20 & 2.99 $\times$ $10^{0}$ & 2.49 $\times$ $10^{0}$ \\
 & -1 & $-$21 & 3.86 $\times$ $10^{0}$ & 3.18 $\times$ $10^{0}$ \\
 & -3 & $-$23 & 6.37 $\times$ $10^{0}$ & 5.22 $\times$ $10^{0}$ \\
 & -5 & $-$25 & 1.05 $\times$ $10^{1}$ & 8.62 $\times$ $10^{0}$ \\
30 & 30 & 0 & 1.05 $\times$ $10^{-2}$ & 1.63 $\times$ $10^{-1}$ \\  
 & 20 & $-$10 & 1.88 $\times$ $10^{-2}$ & 2.37 $\times$ $10^{-1}$ \\ 
 & 15 & $-$15 & 5.09 $\times$ $10^{-2}$ & 3.19 $\times$ $10^{-1}$ \\
 & 10 & $-$20 & 2.82 $\times$ $10^{0}$ & 3.43 $\times$ $10^{0}$ \\
 & 7 & $-$23 & 5.92 $\times$ $10^{0}$ & 4.28 $\times$ $10^{0}$ \\
50 & 50 & 0 & 8.50 $\times$ $10^{-3}$ & 2.23 $\times$ $10^{-1}$ \\  
 & 30 & $-$20 & 2.16 $\times$ $10^{-2}$ & 2.70 $\times$ $10^{-1}$ \\
\end{longtable}

\begin{longtable}{>{\centering\arraybackslash}p{2.5cm}>{\centering\arraybackslash}p{2.5cm}>{\centering\arraybackslash}p{2.5cm}>{\centering\arraybackslash}p{2.5cm}>{\centering\arraybackslash}p{2.5cm}}
\caption{The same as Table I, for \emph{a} = $10^{-5}$.}
\label{tab:longtable5} \\
\hline
\hline
$W_{0}$ & $\theta_{0}$ & $\theta_{R}$ & $\tilde{R}$ & $\tilde{M}$\\ 
\hline
\endfirsthead
\multicolumn{5}{c}{Table V (\emph{continued}).} \\
\hline
\hline
$W_{0}$ & $\theta_{0}$ & $\theta_{R}$ & $\tilde{R}$ & $\tilde{M}$ \\ 
\hline
\endhead
\hline
\endfoot
\hline
\endlastfoot
1 & 1 & 0 & 6.86 $\times$ $10^{-3}$ & 2.85 $\times$ $10^{-3}$ \\ 
 & 0 & $-$1 & 7.92 $\times$ $10^{-3}$ & 3.21 $\times$ $10^{-3}$ \\ 
 & $-$1 & $-$2 & 9.62 $\times$ $10^{-3}$ & 3.83 $\times$ $10^{-3}$ \\
 & $-$3 & $-$4 & 1.54 $\times$ $10^{-2}$ & 6.04 $\times$ $10^{-3}$ \\ 
 & $-$5 & $-$6 & 2.52 $\times$ $10^{-2}$ & 9.92 $\times$ $10^{-3}$ \\ 
 & $-$7 & $-$8 & 4.15 $\times$ $10^{-2}$ & 1.64 $\times$ $10^{-2}$ \\
 & $-$10 & $-$11 & 8.79 $\times$ $10^{-2}$ & 3.43 $\times$ $10^{-2}$ \\
 & $-$15 & $-$16 & 3.07 $\times$ $10^{-1}$ & 1.21 $\times$ $10^{-1}$ \\
 & $-$20 & $-$21 & 1.07 $\times$ $10^{0}$ & 4.22 $\times$ $10^{-1}$ \\
 & $-$30 & $-$31 & 1.31 $\times$ $10^{1}$ & 5.15 $\times$ $10^{0}$ \\
3 & 3 & 0 & 3.48 $\times$ $10^{-3}$ & 4.45 $\times$ $10^{-3}$ \\  
 & 1 & $-$2 & 4.65 $\times$ $10^{-3}$ & 5.21 $\times$ $10^{-3}$ \\  
 & 0 & $-$3 & 5.78 $\times$ $10^{-3}$ & 6.12 $\times$ $10^{-3}$ \\
 & $-$1 & $-$4 & 7.34 $\times$ $10^{-3}$ & 7.53 $\times$ $10^{-3}$ \\
 & $-$3 & $-$6 & 1.21 $\times$ $10^{-2}$ & 1.22 $\times$ $10^{-2}$ \\
 & $-$5 & $-$8 & 1.99 $\times$ $10^{-2}$ & 2.00 $\times$ $10^{-2}$ \\
 & $-$7 & $-$10 & 3.27 $\times$ $10^{-2}$ & 3.31 $\times$ $10^{-2}$ \\
 & $-$-10 & $-$13 & 6.93 $\times$ $10^{-2}$ & 7.02 $\times$ $10^{-2}$ \\
 & $-$15 & $-$18 & 2.42 $\times$ $10^{-1}$ & 2.44 $\times$ $10^{-1}$ \\
 & $-$20 & $-$23 & 8.45 $\times$ $10^{-1}$ & 8.55 $\times$ $10^{-1}$ \\
 & $-$30 & $-$33 & 1.03 $\times$ $10^{1}$ & 1.04 $\times$ $10^{1}$ \\
5 & 5 & 0 & 2.57 $\times$ $10^{-3}$ & 5.69 $\times$ $10^{-3}$ \\  
 & 3 & $-$2 & 3.25 $\times$ $10^{-3}$ & 6.05 $\times$ $10^{-3}$ \\  
 & 1 & $-$4 & 5.05 $\times$ $10^{-3}$ & 7.55 $\times$ $10^{-3}$ \\  
 & 0 & $-$5 & 6.51 $\times$ $10^{-3}$ & 9.11 $\times$ $10^{-3}$ \\
 & $-$1 & $-$6 & 8.40 $\times$ $10^{-3}$ & 1.14 $\times$ $10^{-2}$ \\
 & $-$3 & $-$8 & 1.39 $\times$ $10^{-2}$ & 1.86 $\times$ $10^{-2}$ \\
 & $-$5 & $-$10 & 2.30 $\times$ $10^{-2}$ & 3.07 $\times$ $10^{-2}$ \\
 & $-$7 & $-$12 & 3.79 $\times$ $10^{-2}$ & 5.07 $\times$ $10^{-2}$ \\
 & $-$10 & $-$15 & 8.02 $\times$ $10^{-2}$ & 1.08 $\times$ $10^{-1}$ \\
 & $-$15 & $-$20 & 2.80 $\times$ $10^{-1}$ & 3.75 $\times$ $10^{-1}$ \\
 & $-$20 & $-$25 & 9.77 $\times$ $10^{-1}$ & 1.31 $\times$ $10^{0}$ \\
 & $-$30 & $-$35 & 1.19 $\times$ $10^{1}$ & 1.60 $\times$ $10^{1}$ \\
7 & 7 & 0 & 2.12 $\times$ $10^{-3}$ & 6.82 $\times$ $10^{-3}$ \\  
 & 5 & $-$2 & 2.56 $\times$ $10^{-3}$ & 6.90 $\times$ $10^{-3}$ \\  
 & 3 & $-$4 & 3.80 $\times$ $10^{-3}$ & 7.56 $\times$ $10^{-3}$ \\  
 & 1 & $-$6 & 6.55 $\times$ $10^{-3}$ & 9.91 $\times$ $10^{-3}$ \\ 
 & 0 & $-$7 & 8.63 $\times$ $10^{-3}$ & 1.22 $\times$ $10^{-2}$ \\
 & $-$1 & $-$8 & 1.12 $\times$ $10^{-2}$ & 1.54 $\times$ $10^{-2}$ \\
 & $-$3 & $-$10 & 1.87 $\times$ $10^{-2}$ & 2.52 $\times$ $10^{-2}$ \\
 & $-$5 & $-$12 & 3.09 $\times$ $10^{-2}$ & 4.17 $\times$ $10^{-2}$ \\
 & $-$7 & $-$14 & 5.11 $\times$ $10^{-2}$ & 6.90 $\times$ $10^{-2}$ \\
 & $-$10 & $-$17 & 1.08 $\times$ $10^{-1}$ & 1.45 $\times$ $10^{-1}$ \\
 & $-$15 & $-$22 & 3.77 $\times$ $10^{-1}$ & 5.10 $\times$ $10^{-1}$ \\
 & $-$20 & $-$27 & 1.32 $\times$ $10^{0}$ & 1.78 $\times$ $10^{0}$ \\
 & $-$30 & $-$37 & 1.61 $\times$ $10^{1}$ & 2.17 $\times$ $10^{1}$ \\
10 & 10 & 0 & 1.76 $\times$ $10^{-3}$ & 8.37 $\times$ $10^{-3}$ \\  
 & 7 & $-$3 & 2.29 $\times$ $10^{-3}$ & 8.40 $\times$ $10^{-3}$ \\  
 & 5 & $-$5 & 3.41 $\times$ $10^{-3}$ & 8.83 $\times$ $10^{-3}$ \\  
 & 3 & $-$7 & 6.38 $\times$ $10^{-3}$ & 9.53 $\times$ $10^{-3}$ \\  
 & 1 & $-$9 & 1.31 $\times$ $10^{-2}$ & 1.35 $\times$ $10^{-2}$ \\  
 & 0 & $-$10 & 1.80 $\times$ $10^{-2}$ & 1.71 $\times$ $10^{-2}$ \\
 & $-$1 & $-$11 & 2.37 $\times$ $10^{-2}$ & 2.01 $\times$ $10^{-2}$ \\
 & $-$3 & $-$13 & 3.98 $\times$ $10^{-2}$ & 3.45 $\times$ $10^{-2}$ \\
 & $-$5 & $-$15 & 6.58 $\times$ $10^{-2}$ & 5.83 $\times$ $10^{-2}$ \\
 & $-$7 & $-$17 & 1.09 $\times$ $10^{-1}$ & 9.76 $\times$ $10^{-2}$ \\
 & $-$10 & $-$20 & 2.30 $\times$ $10^{-1}$ & 2.09 $\times$ $10^{-1}$ \\
 & $-$15 & $-$25 & 8.03 $\times$ $10^{-1}$ & 7.36 $\times$ $10^{-1}$ \\
 & $-$-20 & $-$30 & 2.80 $\times$ $10^{0}$ & 2.57 $\times$ $10^{0}$ \\
 & $-$30 & $-$40 & 3.41 $\times$ $10^{1}$ & 3.14 $\times$ $10^{1}$ \\
15 & 15 & 0 & 1.44 $\times$ $10^{-3}$ & 1.07 $\times$ $10^{-2}$ \\  
 & 10 & $-$5 & 2.18 $\times$ $10^{-3}$ & 1.37 $\times$ $10^{-2}$ \\
 & 7 & $-$-8 & 4.48 $\times$ $10^{-3}$ & 1.87 $\times$ $10^{-2}$ \\
 & 5 & $-$10 & 1.39 $\times$ $10^{-2}$ & 2.00 $\times$ $10^{-2}$ \\
 & 3 & $-$12 & 7.37 $\times$ $10^{-2}$ & 2.22 $\times$ $10^{-2}$ \\
 & 1 & $-$14 & 9.75 $\times$ $10^{-2}$ & 4.84 $\times$ $10^{-2}$ \\
 & 0 & $-$15 & 1.19 $\times$ $10^{-1}$ & 6.23 $\times$ $10^{-2}$ \\
 & $-$1 & $-$16 & 1.51 $\times$ $10^{-1}$ & 7.95 $\times$ $10^{-2}$ \\
 & $-$3 & $-$18 & 2.47 $\times$ $10^{-1}$ & 1.33 $\times$ $10^{-1}$ \\
 & $-$5 & $-$20 & 4.06 $\times$ $10^{-1}$ & 2.22 $\times$ $10^{-1}$ \\
 & $-$7 & $-$22 & 6.70 $\times$ $10^{-1}$ & 3.68 $\times$ $10^{-1}$ \\
 & $-$10 & $-$25 & 1.42 $\times$ $10^{0}$ & 7.83 $\times$ $10^{-1}$ \\
 & $-$15 & $-$30 & 4.95 $\times$ $10^{0}$ & 2.74 $\times$ $10^{0}$ \\
 & $-$20 & $-$35 & 1.73 $\times$ $10^{1}$ & 9.56 $\times$ $10^{0}$ \\
20 & 20 & 0 & 1.26 $\times$ $10^{-3}$ & 1.28 $\times$ $10^{-2}$ \\  
 & 15 & $-$5 & 1.70 $\times$ $10^{-3}$ & 1.58 $\times$ $10^{-2}$ \\
 & 10 & $-$10 & 4.35 $\times$ $10^{-3}$ & 2.02 $\times$ $10^{-2}$ \\
 & 7 & $-$13 & 1.96 $\times$ $10^{-1}$ & 2.19 $\times$ $10^{-2}$ \\
 & 5 & $-$15 & 2.07 $\times$ $10^{-1}$ & 8.32 $\times$ $10^{-2}$ \\
 & 3 & $-$17 & 2.13 $\times$ $10^{-1}$ & 1.26 $\times$ $10^{-1}$ \\
 & 1 & $-$19 & 2.32 $\times$ $10^{-1}$ & 1.94 $\times$ $10^{-1}$ \\
 & 0 & $-$20 & 2.99 $\times$ $10^{-1}$ & 2.46 $\times$ $10^{-1}$ \\
 & $-$1 & $-$21 & 3.85 $\times$ $10^{-1}$ & 3.15 $\times$ $10^{-1}$ \\
 & $-$3 & $-$23 & 6.37 $\times$ $10^{-1}$ & 5.20 $\times$ $10^{-1}$ \\
 & $-$5 & $-$25 & 1.05 $\times$ $10^{0}$ & 8.60 $\times$ $10^{-1}$ \\
 & $-$7 & $-$27 & 1.73 $\times$ $10^{0}$ & 1.42 $\times$ $10^{0}$ \\
 & $-$10 & $-$30 & 3.68 $\times$ $10^{0}$ & 3.00 $\times$ $10^{0}$ \\
 & $-$15 & $-$35 & 1.28 $\times$ $10^{1}$ & 1.05 $\times$ $10^{1}$ \\
30 & 30 & 0 & 1.05 $\times$ $10^{-3}$ & 1.64 $\times$ $10^{-2}$ \\  
 & 20 & $-$10 & 1.88 $\times$ $10^{-3}$ & 2.37 $\times$ $10^{-2}$ \\ 
 & 15 & $-$15 & 5.06 $\times$ $10^{-3}$ & 3.19 $\times$ $10^{-2}$ \\
 & 10 & $-$20 & 2.83 $\times$ $10^{-1}$ & 3.42 $\times$ $10^{-1}$ \\
 & 7 & $-$23 & 5.88 $\times$ $10^{-1}$ & 4.26 $\times$ $10^{-1}$ \\
 & 5 & $-$25 & 1.28 $\times$ $10^{0}$ & 7.31 $\times$ $10^{-1}$ \\
 & 3 & $-$27 & 1.99 $\times$ $10^{0}$ & 1.33 $\times$ $10^{0}$ \\
 & 1 & $-$29 & 3.26 $\times$ $10^{0}$ & 2.25 $\times$ $10^{0}$ \\
 & 0 & $-$30 & 4.03 $\times$ $10^{0}$ & 2.85 $\times$ $10^{0}$ \\
 & $-$1 & $-$31 & 5.19 $\times$ $10^{0}$ & 3.65 $\times$ $10^{0}$ \\
 & $-$3 & $-$33 & 8.57 $\times$ $10^{0}$ & 6.02 $\times$ $10^{0}$ \\
 & $-$5 & $-$35 & 1.41 $\times$ $10^{1}$ & 9.94 $\times$ $10^{0}$ \\
50 & 50 & 0 & 8.50 $\times$ $10^{-4}$ & 2.34 $\times$ $10^{-2}$ \\  
 & 30 & $-$20 & 2.15 $\times$ $10^{-3}$ & 2.71 $\times$ $10^{-2}$ \\
 & 20 & $-$30 & 4.08 $\times$ $10^{0}$ & 4.83 $\times$ $10^{0}$ \\
\end{longtable}

\begin{table}
\caption{Lower limits on the value of masses of particles obtained by Eq.(\ref{massa_3}), for $\rho(r_{c}) = 9.38 \times 10^{-30}$ g/cm$^{3}$ and $\sigma = 100$ km/s. The values of masses are e$\times$pressed in eV.}
\label{tab:table6}
\begin{center}
\begin{tabular}{>{\centering\arraybackslash}p{2.5cm}>{\centering\arraybackslash}p{2.5cm}>{\centering\arraybackslash}p{2.5cm}>{\centering\arraybackslash}p{2.5cm}>{\centering\arraybackslash}p{2.5cm}>{\centering\arraybackslash}p{2.5cm}}
\hline
\hline
$W_{0}$ & $a = 10^{-5}$ & $a = 10^{-3}$ & $a = 10^{-1}$ & $a = 0.5$ & $a = 1$ \\ 
\hline
10 & 9.58 $\times$ $10^{-2}$ & 2.98 $\times$ $10^{-1}$ & 9.31 $\times$ $10^{-1}$ & 1.12 $\times$ $10^{0}$ & 1.13 $\times$ $10^{0}$ \\ 
15 & 7.99 $\times$ $10^{-2}$ & 2.53 $\times$ $10^{-1}$ & 7.95 $\times$ $10^{-1}$ & 9.66 $\times$ $10^{-1}$ & 9.78 $\times$ $10^{-1}$ \\ 
20 & 7.21 $\times$ $10^{-2}$ & 2.25 $\times$ $10^{-1}$ & 7.09 $\times$ $10^{-1}$ & 8.68 $\times$ $10^{-1}$ & 8.80 $\times$ $10^{-1}$ \\ 
30 & 6.03 $\times$ $10^{-2}$ & 1.91 $\times$ $10^{-1}$ & 6.01 $\times$ $10^{-1}$ & 7.45 $\times$ $10^{-1}$ & 7.57 $\times$ $10^{-1}$ \\ 
50 & 4.86 $\times$ $10^{-2}$ & 1.54 $\times$ $10^{-1}$ & 4.87 $\times$ $10^{-1}$ & 6.13 $\times$ $10^{-1}$ & 6.26 $\times$ $10^{-1}$ \\ 
\hline
\end{tabular}
\end{center}
\end{table}

\begin{table}
\caption{The same as Table VI, for $\rho(r_{c}) = 9.38$ $\times$ $10^{-27}$ $g/cm^{3}$ and $\sigma = 100$ $km/s$.}
\label{tab:table7}
\begin{center}
\begin{tabular}{>{\centering\arraybackslash}p{2.5cm}>{\centering\arraybackslash}p{2.5cm}>{\centering\arraybackslash}p{2.5cm}>{\centering\arraybackslash}p{2.5cm}>{\centering\arraybackslash}p{2.5cm}>{\centering\arraybackslash}p{2.5cm}}
\hline
\hline
$W_{0}$ & $a = 10^{-5}$ & $a = 10^{-3}$ & $a = 10^{-1}$ & $a = 0.5$ & $a = 1$ \\ 
\hline
10 & 5.38 $\times$ $10^{-1}$ & 1.68 $\times$ $10^{0}$ & 5.23 $\times$ $10^{0}$ & 6.30 $\times$ $10^{0}$ & 6.36 $\times$ $10^{0}$ \\ 
15 & 4.50 $\times$ $10^{-1}$ & 1.42 $\times$ $10^{0}$ & 4.47 $\times$ $10^{0}$ & 5.41 $\times$ $10^{0}$ & 5.51 $\times$ $10^{0}$ \\ 
20 & 4.06 $\times$ $10^{-1}$ & 1.26 $\times$ $10^{0}$ & 4.00 $\times$ $10^{0}$ & 4.88 $\times$ $10^{0}$ & 4.94 $\times$ $10^{0}$ \\ 
30 & 3.40 $\times$ $10^{-1}$ & 1.07 $\times$ $10^{0}$ & 3.37 $\times$ $10^{0}$ & 4.19 $\times$ $10^{0}$ & 4.25 $\times$ $10^{0}$ \\ 
50 & 2.73 $\times$ $10^{-1}$ & 8.67 $\times$ $10^{-1}$ & 2.74 $\times$ $10^{0}$ & 3.46 $\times$ $10^{0}$ & 3.53 $\times$ $10^{0}$ \\ 
\hline
\end{tabular}
\end{center}
\end{table}

\begin{table}
\caption{The same as Table VI, for $\rho(r_{c}) = 9.38 \times 10^{-24}$ g/cm$^{3}$ and $\sigma = 100$ km/s.}
\label{tab:table8}
\begin{center}
\begin{tabular}{>{\centering\arraybackslash}p{2.5cm}>{\centering\arraybackslash}p{2.5cm}>{\centering\arraybackslash}p{2.5cm}>{\centering\arraybackslash}p{2.5cm}>{\centering\arraybackslash}p{2.5cm}>{\centering\arraybackslash}p{2.5cm}}
\hline
\hline
$W_{0}$ & $a = 10^{-5}$ & $a = 10^{-3}$ & $a = 10^{-1}$ & $a = 0.5$ & $a = 1$ \\ 
\hline
10 & 3.03 $\times$ $10^{0}$ & 9.44 $\times$ $10^{0}$ & 2.94 $\times$ $10^{1}$ & 3.54 $\times$ $10^{1}$ & 3.58 $\times$ $10^{1}$ \\ 
15 & 2.53 $\times$ $10^{0}$ & 7.99 $\times$ $10^{0}$ & 2.51 $\times$ $10^{1}$ & 3.05 $\times$ $10^{1}$ & 3.10 $\times$ $10^{1}$ \\ 
20 & 2.28 $\times$ $10^{0}$ & 7.10 $\times$ $10^{0}$ & 2.25 $\times$ $10^{1}$ & 2.75 $\times$ $10^{1}$ & 2.78 $\times$ $10^{1}$ \\ 
30 & 1.91 $\times$ $10^{0}$ & 6.04 $\times$ $10^{0}$ & 1.89 $\times$ $10^{1}$ & 2.36 $\times$ $10^{1}$ & 2.39 $\times$ $10^{1}$ \\ 
50 & 1.54 $\times$ $10^{0}$ & 4.87 $\times$ $10^{0}$ & 1.54 $\times$ $10^{1}$ & 1.95 $\times$ $10^{1}$ & 1.98 $\times$ $10^{1}$ \\ 
\hline
\end{tabular}
\end{center}
\end{table}

\begin{table}
\caption{The same as Table VI, for $\rho(r_{c}) = 9.38 \times 10^{-21}$ g/cm$^{3}$ and $\sigma = 100$ km/s.}
\label{tab:table9}
\begin{center}
\begin{tabular}{>{\centering\arraybackslash}p{2.5cm}>{\centering\arraybackslash}p{2.5cm}>{\centering\arraybackslash}p{2.5cm}>{\centering\arraybackslash}p{2.5cm}>{\centering\arraybackslash}p{2.5cm}>{\centering\arraybackslash}p{2.5cm}}
\hline
\hline
$W_{0}$ & $a = 10^{-5}$ & $a = 10^{-3}$ & $a = 10^{-1}$ & $a = 0.5$ & $a = 1$ \\ 
\hline 
10 & 1.70 $\times$ $10^{1}$ & 5.31 $\times$ $10^{1}$ & 1.65 $\times$ $10^{2}$ & 1.99 $\times$ $10^{2}$ & 2.01 $\times$ $10^{2}$ \\ 
15 & 1.42 $\times$ $10^{1}$ & 4.49 $\times$ $10^{1}$ & 1.41 $\times$ $10^{2}$ & 1.71 $\times$ $10^{2}$ & 1.74 $\times$ $10^{2}$ \\ 
20 & 1.28 $\times$ $10^{2}$ & 3.99 $\times$ $10^{1}$ & 1.26 $\times$ $10^{2}$ & 1.54 $\times$ $10^{2}$ & 1.56 $\times$ $10^{2}$ \\ 
30 & 1.08 $\times$ $10^{1}$ & 3.40 $\times$ $10^{1}$ & 1.07 $\times$ $10^{2}$ & 1.32 $\times$ $10^{2}$ & 1.34 $\times$ $10^{2}$ \\ 
50 & 8.63 $\times$ $10^{0}$ & 2.74 $\times$ $10^{1}$ & 8.65 $\times$ $10^{1}$ & 1.09 $\times$ $10^{2}$ & 1.11 $\times$ $10^{2}$ \\ 
\hline
\end{tabular}
\end{center}
\end{table}

\begin{table}
\caption{The same as Table VI, for $\rho(r_{c}) = 9.38 \times 10^{-18}$ g/cm$^{3}$ and $\sigma = 100$ km/s.}
\label{tab:table10}
\begin{center}
\begin{tabular}{>{\centering\arraybackslash}p{2.5cm}>{\centering\arraybackslash}p{2.5cm}>{\centering\arraybackslash}p{2.5cm}>{\centering\arraybackslash}p{2.5cm}>{\centering\arraybackslash}p{2.5cm}>{\centering\arraybackslash}p{2.5cm}}
\hline
\hline
$W_{0}$ & $a = 10^{-5}$ & $a = 10^{-3}$ & $a = 10^{-1}$ & $a = 0.5$ & $a = 1$ \\ 
\hline
10 & 9.58 $\times$ $10^{1}$ & 2.99 $\times$ $10^{2}$ & 9.30 $\times$ $10^{2}$ & 1.12 $\times$ $10^{3}$ & 1.13 $\times$ $10^{3}$\\ 
15 & 8.01 $\times$ $10^{1}$ & 2.53 $\times$ $10^{2}$ & 7.95 $\times$ $10^{2}$ & 9.63 $\times$ $10^{2}$ & 9.80 $\times$ $10^{2}$ \\ 
20 & 7.22 $\times$ $10^{1}$ & 2.25 $\times$ $10^{2}$ & 7.11 $\times$ $10^{2}$ & 8.68 $\times$ $10^{2}$ & 8.79 $\times$ $10^{2}$ \\ 
30 & 6.05 $\times$ $10^{1}$ & 1.91 $\times$ $10^{2}$ & 5.99 $\times$ $10^{2}$ & 7.45 $\times$ $10^{2}$ & 7.56 $\times$ $10^{2}$ \\ 
50 & 4.86 $\times$ $10^{1}$ & 1.54 $\times$ $10^{2}$ & 4.87 $\times$ $10^{2}$ & 6.16 $\times$ $10^{2}$ & 6.27 $\times$ $10^{2}$ \\ 
\hline
\end{tabular}
\end{center}
\end{table}

\begin{table}
\caption{Values of $\theta_{0}$ for $m_* = 3$ keV.}
\label{tab:table11}
\begin{center}
\begin{tabular}{>{\centering\arraybackslash}p{2.5cm}>{\centering\arraybackslash}p{2.5cm}>{\centering\arraybackslash}p{2.5cm}>{\centering\arraybackslash}p{2.5cm}>{\centering\arraybackslash}p{2.5cm}>{\centering\arraybackslash}p{2.5cm}}
\hline
\hline
\multicolumn{2}{c}{} & \multicolumn{2}{c}{$\rho = 9.38 \times 10^{-20}$ g/cm$^{3}$} & \multicolumn{2}{c}{$\rho = 9.38 \times 10^{-25}$ g/cm$^{3}$} \\
\hline
$W_{0}$ & $\sigma$ (km/s) & $\theta_{0}$ (semideg.) & $\theta_{0}$ (classical) & $\theta_{0}$ (semideg.) & $\theta_{0}$ (classical) \\ 
\hline
5 & 250 & $-$8.48 & $-$8.48 & $-$19.9 & $-$19.9 \\ 
 & 150 & $-$6.95 & $-$6.95 & $-$18.5 & $-$18.5 \\ 
 & 100 & $-$5.73 & $-$5.73 & $-$17.2 & $-$17.2 \\ 
 & 50 & $-$3.64 & $-$3.65 & $-$15.2 & $-$15.2 \\ 
 & 10 & 2.52 & 1.18 & $-$10.3 & $-$10.3 \\ 
3 & 250 & $-$8.19 & $-$8.19 & $-$19.7 & $-$19.7 \\ 
 & 150 & $-$6.66 & $-$6.66 & $-$18.2 & $-$18.2 \\ 
 & 100 & $-$5.44 & $-$5.44 & $-$17.0 & $-$17.0 \\ 
 & 50 & $-$3.35 & $-$3.37 & $-$14.9 & $-$14.9 \\ 
 & 10 & -- & 1.46 & $-$10.0 & $-$10.0 \\ 
1 & 250 & $-$6.67 & $-$6.67 & $-$18.2 & $-$18.2 \\ 
 & 150 & $-$5.13 & $-$5.14 & $-$16.7 & $-$16.7 \\ 
 & 100 & $-$3.90 & $-$3.92 & $-$15.4 & $-$15.4 \\ 
 & 50 & $-$1.72 & $-$1.84 & $-$13.4 & $-$13.4 \\ 
 & 10 & -- & 2.99 & $-$8.52 & $-$8.52 \\
\hline
\end{tabular}
\end{center}
\end{table}

\clearpage

\begin{figure}
\includegraphics[width=12cm]{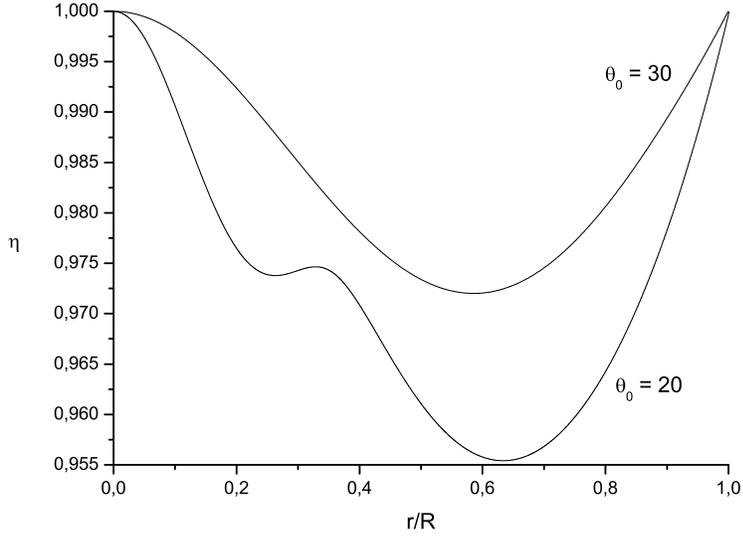}
\caption{Values of the ratio of velocities $\eta$ as a function of the relative radius $r/R$ for the anisotropy parameter \emph{a} = 1, $W_{0}$ = 30, $\theta_{0}$ = 30, and 20.}
\label{fig1}
\end{figure}

\begin{figure}
\includegraphics[width=12cm]{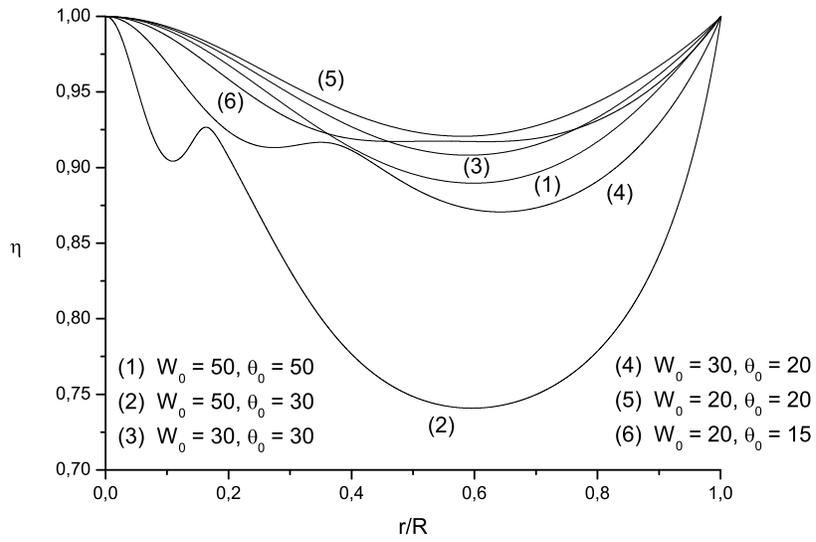}
\caption{The same as in Fig. 1, for \emph{a} = 0.5, $W_{0}$ = 50, and $\theta_{0}$ = 50 and 30; $W_{0}$ = 30, and $\theta_{0}$ = 30 and 20; $W_{0}$ = 20, and $\theta_{0}$ = 20 and 15.}
\label{fig2}
\end{figure}

\begin{figure}
\includegraphics[width=12cm]{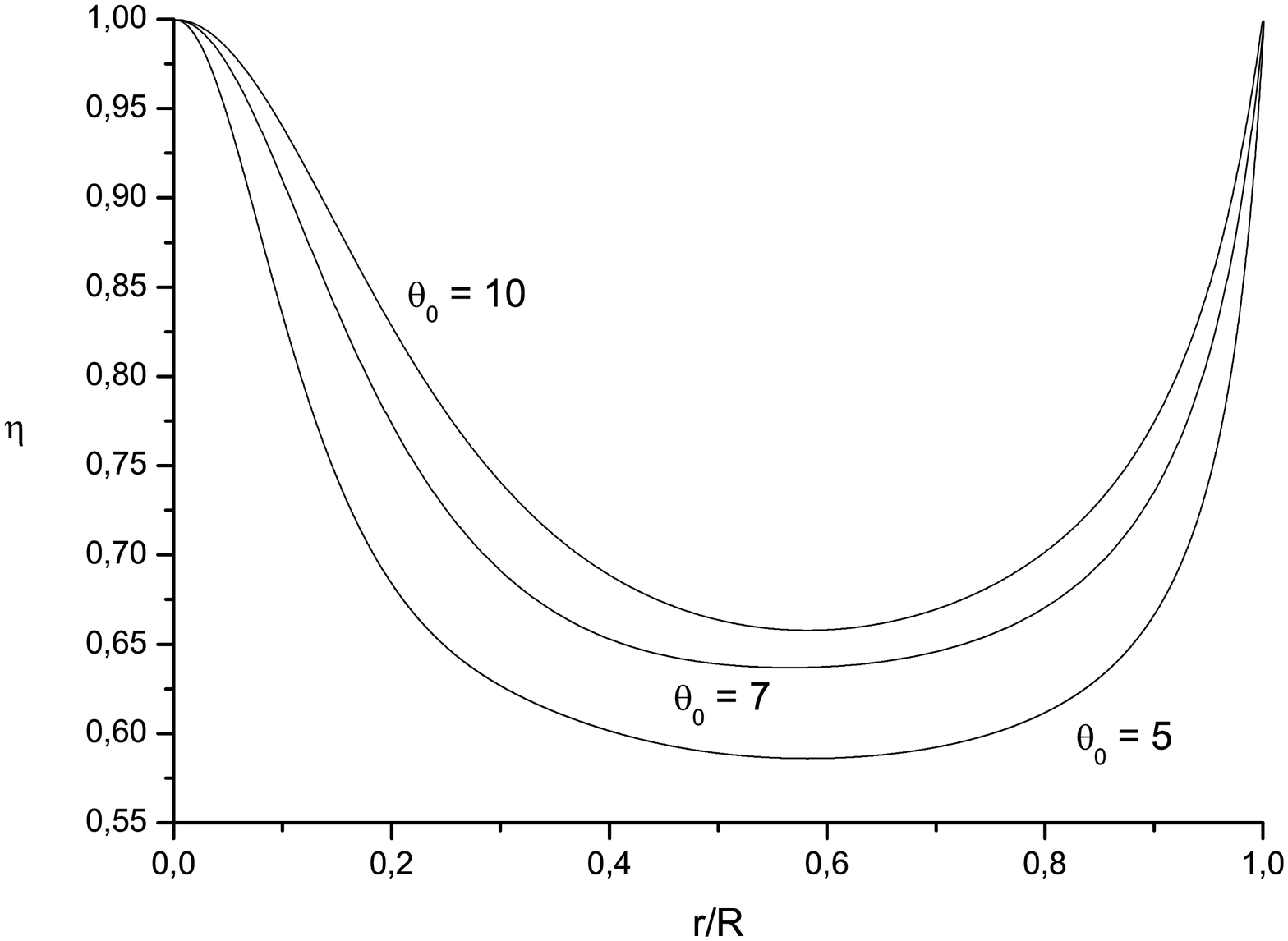}
\caption{The same as in Fig. 1, for \emph{a} = $10^{-1}$, $W_{0}$ = 10, and $\theta_{0}$ = 10, 7, and 5.}
\label{fig3}
\end{figure}

\begin{figure}
\includegraphics[width=12cm]{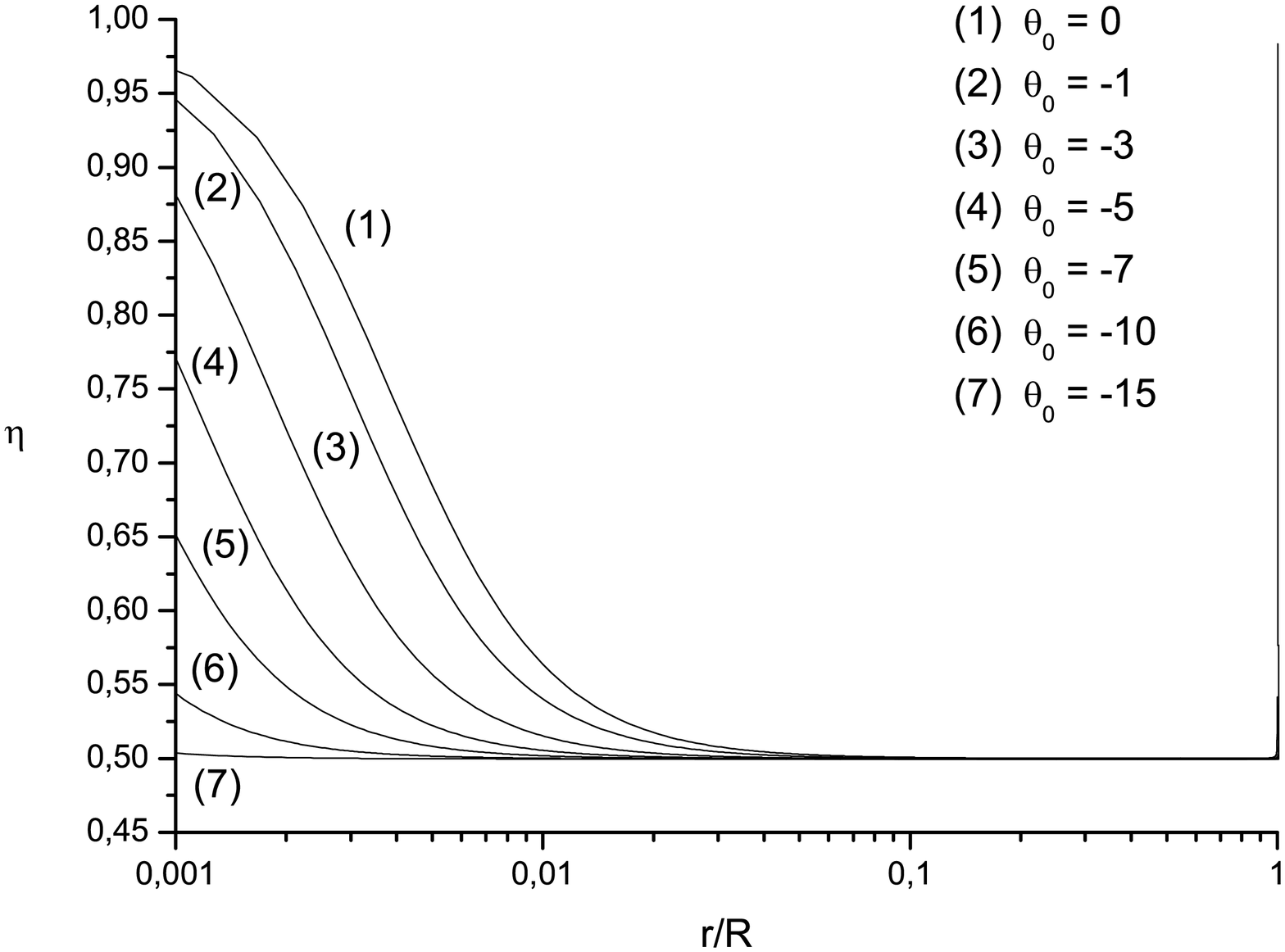}
\caption{The same as in Fig. 1, for \emph{a} = $10^{-3}$, $W_{0}$ = 10, and $\theta_{0} = 0$, $-$1, $-$3, $-$5, $-$7, $-$10, and $-$15.}
\label{fig4}
\end{figure}

\begin{figure}
\includegraphics[width=12cm]{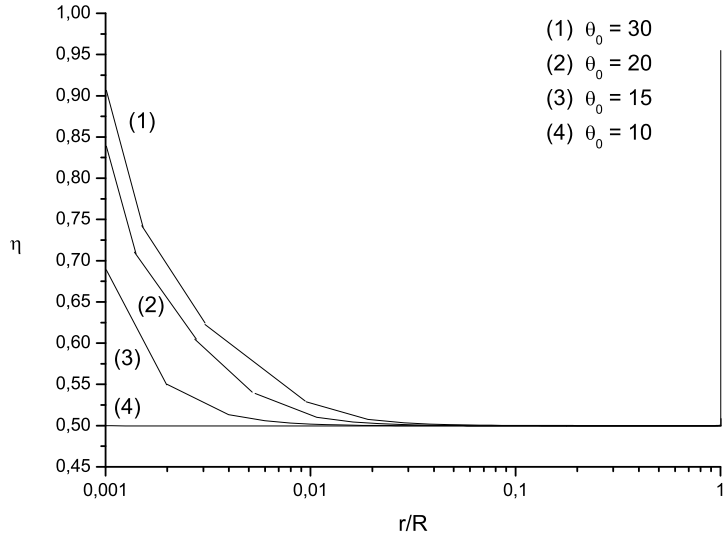}
\caption{The same as in Fig. 1, for \emph{a} = $10^{-5}$, $W_{0}$ = 30, and $\theta_{0}$ = 30, 20, 15, and 10.}
\label{fig5}
\end{figure}

\begin{figure}
\includegraphics[width=12cm]{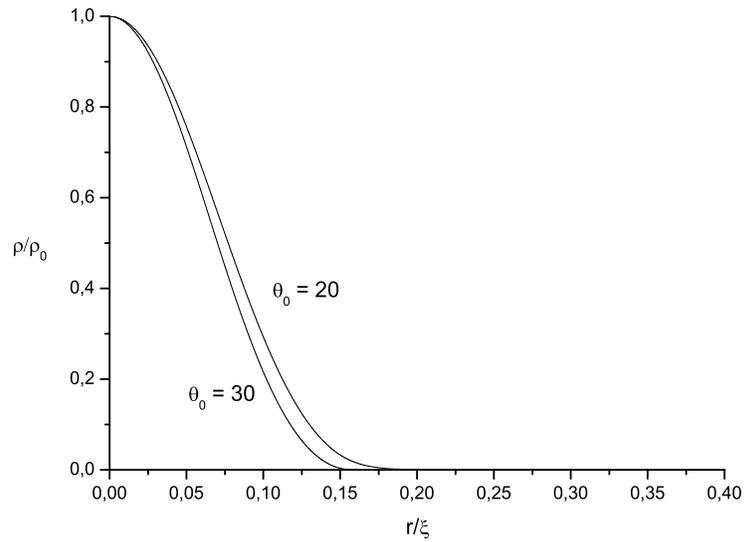}
\caption{Relative density $\rho/\rho_{0}$ as a function of the dimensionless radial coordinate $r/\xi$ for the anisotropy parameter \emph{a} = 1, $W_{0}$ = 30, and $\theta_{0}$ = 30 and 20.}
\label{fig6}
\end{figure}

\begin{figure}
\includegraphics[width=12cm]{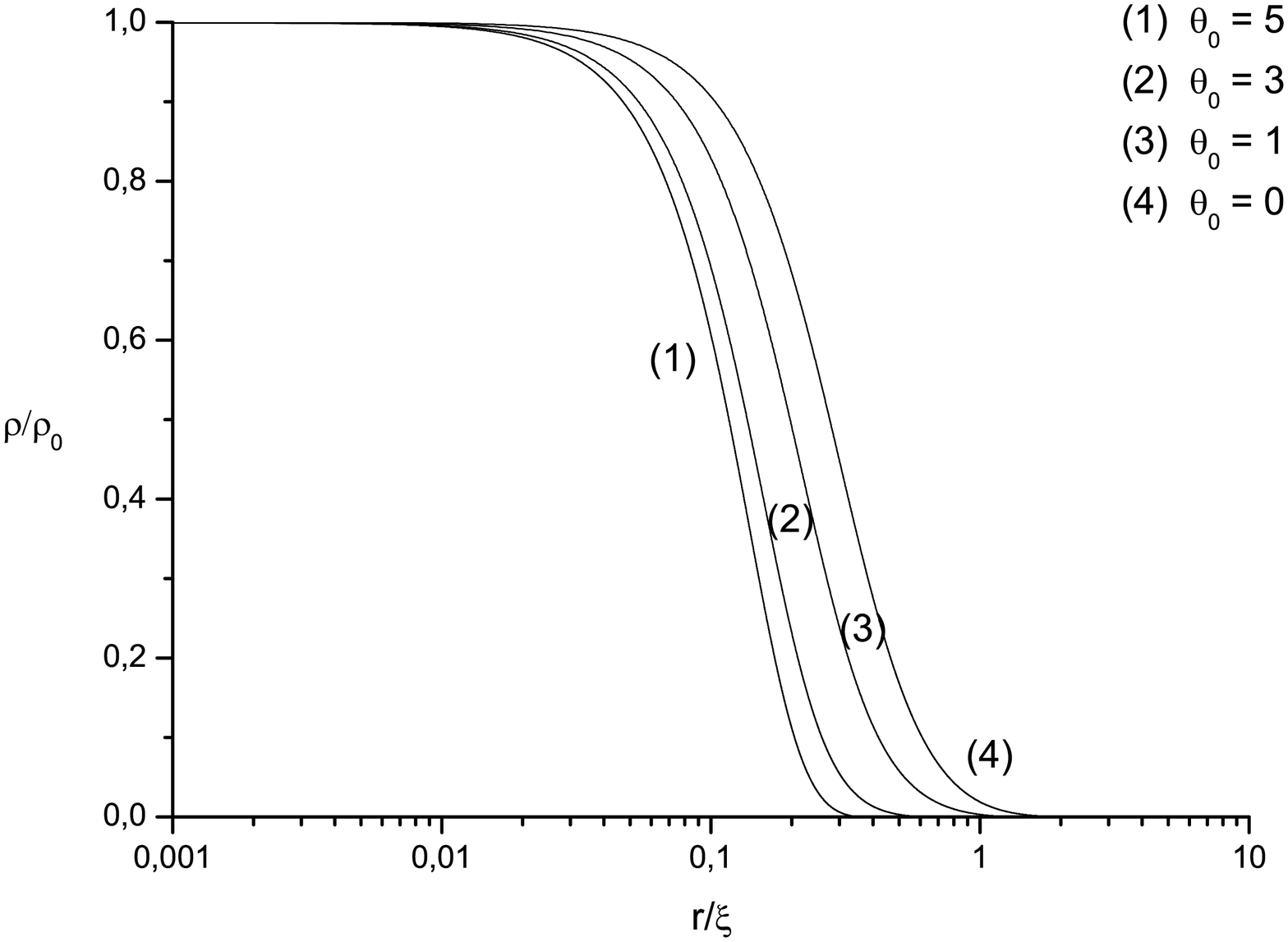}
\caption{The same as in Fig. 6, for \emph{a} = 0.5, $W_{0}$ = 7, and $\theta_{0}$ = 5, 3, 1, and 0.}
\label{fig7}
\end{figure}

\begin{figure}
\includegraphics[width=12cm]{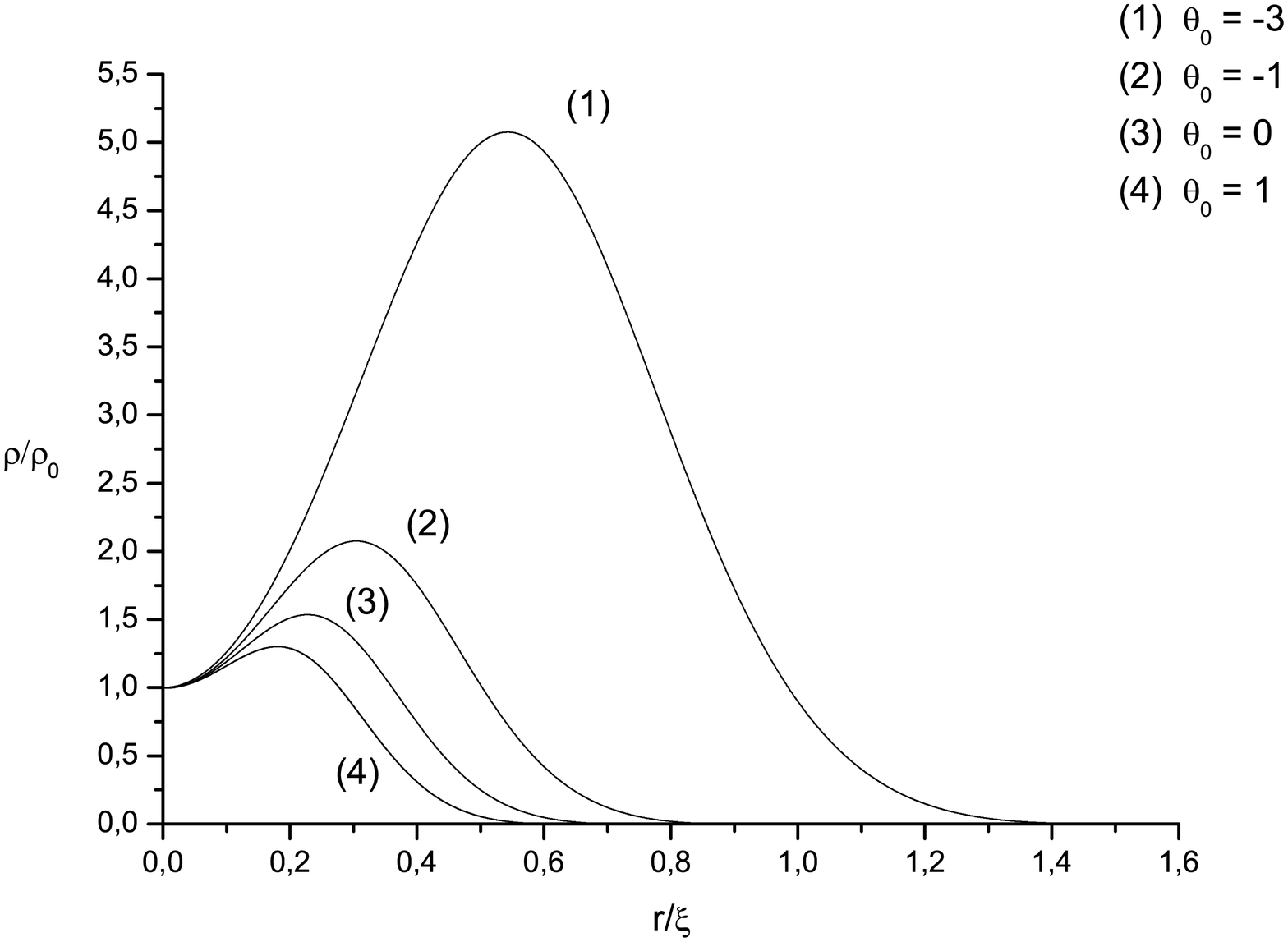}
\caption{The same as in Fig. 6, for \emph{a} = $10^{-1}$, $W_{0}$ = 1, $\theta_{0}$ = 1, 0, $-$1, and $-$3.}
\label{fig8}
\end{figure}

\begin{figure}
\includegraphics[width=12cm]{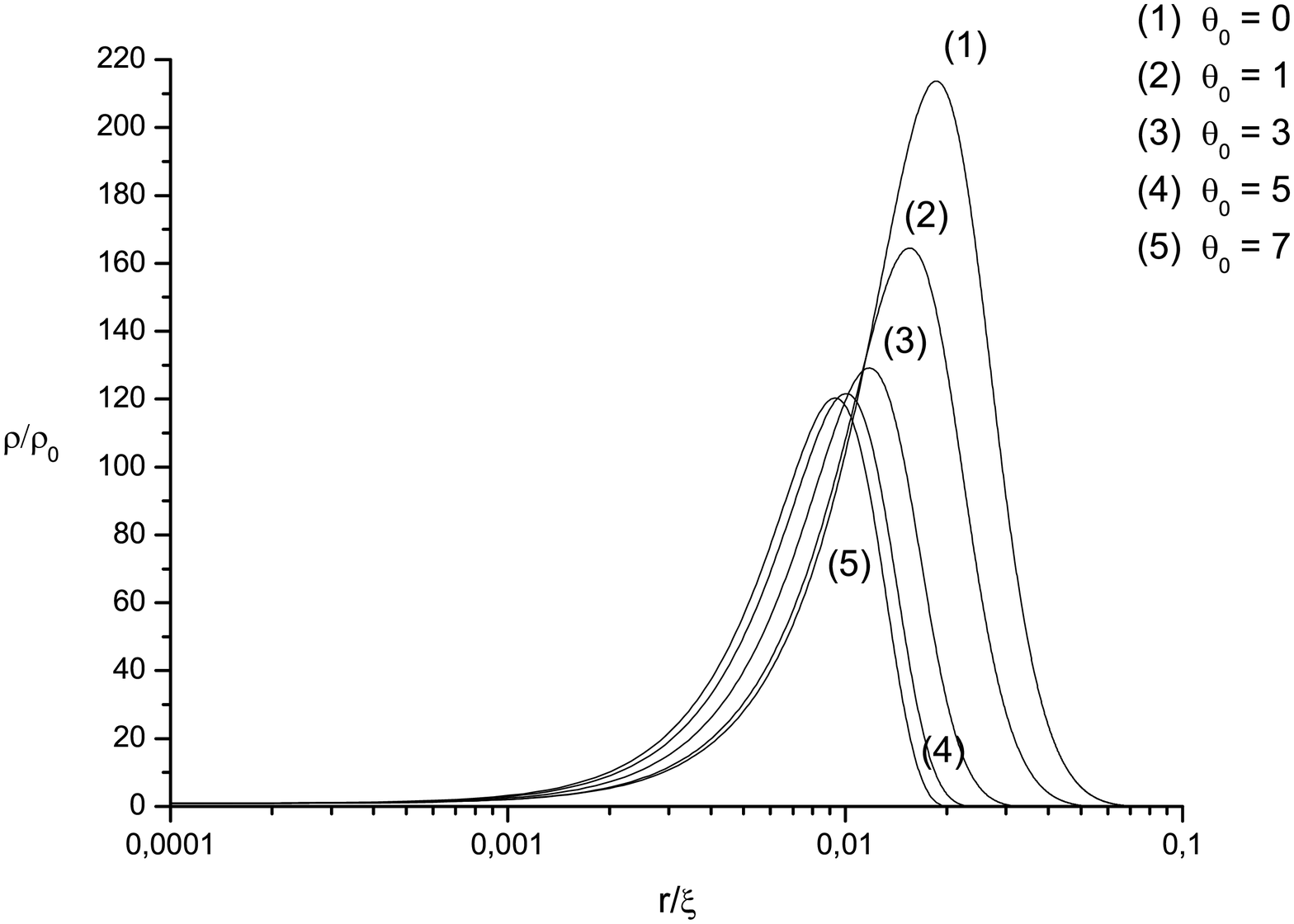}
\caption{The same as in Fig. 6, for \emph{a} = $10^{-3}$, $W_{0}$ = 7, $\theta_{0}$ = 7, 5, 3, 1, and 0.}
\label{fig9}
\end{figure}

\begin{figure}
\includegraphics[width=12cm]{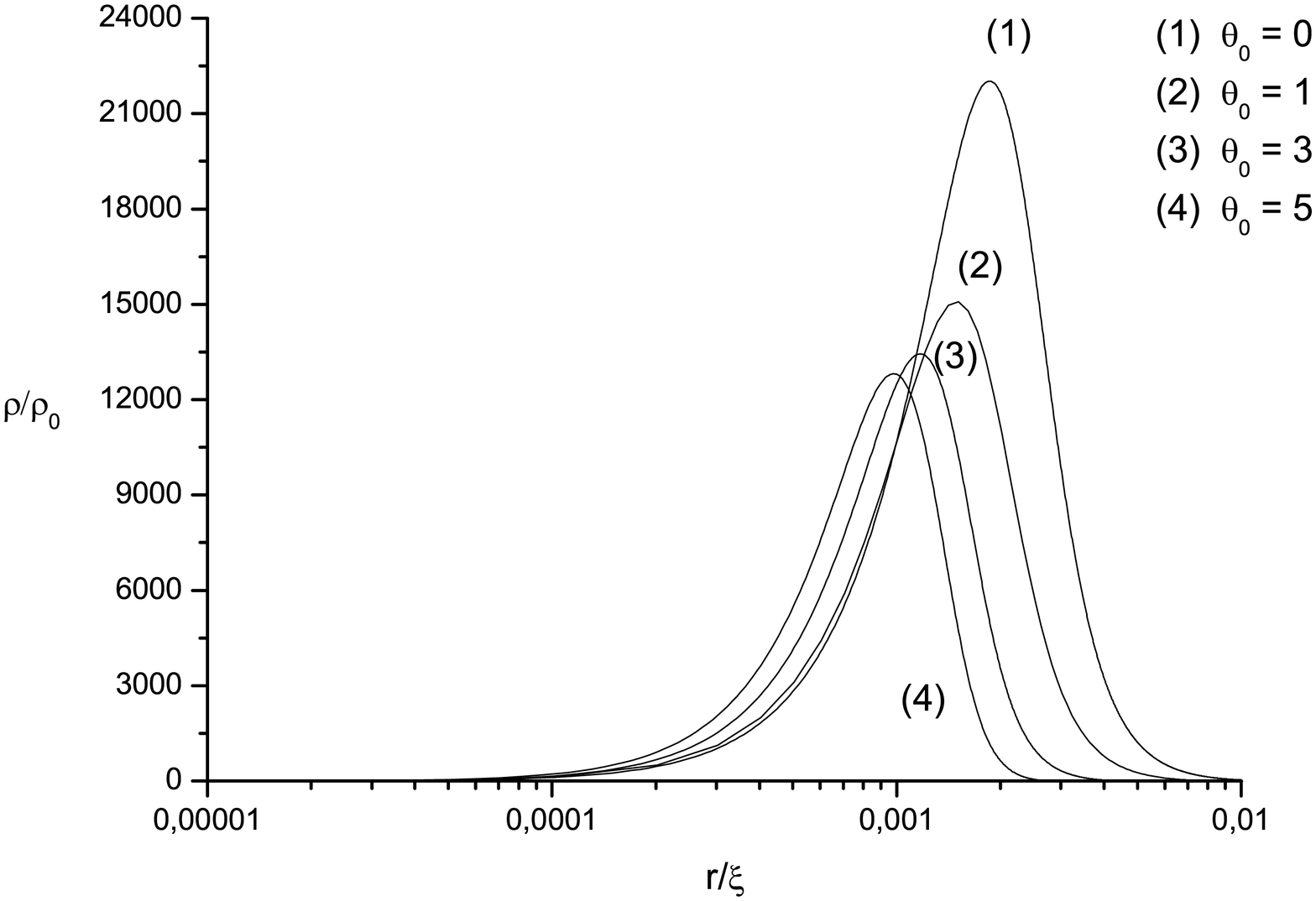}
\caption{The same as in Fig. 6, for \emph{a} = $10^{-5}$, $W_{0}$ = 10, $\theta_{0}$ = 5, 3, 1, and 0.}
\label{fig10}
\end{figure}

\end{document}